\documentclass[aip, amsmath,amssymb, reprint]{revtex4-1}

\draft 
\usepackage{graphicx}
\usepackage{multirow}
\usepackage{mathtools}
\usepackage{hyperref}
\usepackage{xcolor}
\usepackage{array}
\DeclarePairedDelimiterX{\infdivx}[2]{(}{)}{%
  #1\;\delimsize\|\;#2%
}
\newcommand{\KL}{\text{KL}\infdivx}
\begin{document}


\title{On learning latent dynamics of the AUG plasma state} 


\author{A. Kit}%
\email{adam.kit@helsinki.fi}
\affiliation{%
University of Helsinki, FI-00014 Helsinki, Finland
}%
\affiliation{%
VTT Technical Research Centre of Finland, FI-02044 VTT,
Finland
}%
\author{A.E. Järvinen}
\affiliation{%
VTT Technical Research Centre of Finland, FI-02044 VTT,
Finland
}%
\author{Y.R.J. Poels}%
\affiliation{%
Eindhoven University of Technology, Mathematics and Computer Science, NL-5600MB Eindhoven, The Netherlands
}%
\affiliation{%
École Polytechnique Fédérale de Lausanne, Swiss Plasma Center, CH-1015 Lausanne, Switzerland
}%

\author{S. Wiesen}%
\affiliation{%
Forschungszentrum Jülich GmbH, Institut für Energie- und Klimaforschung - Plasmaphysik, DE-52425 Jülich, Germany
}%
\author{V. Menkovski}%
\affiliation{%
Eindhoven University of Technology, Mathematics and Computer Science, NL-5600MB Eindhoven, The Netherlands
}%
\author{R. Fischer}
\affiliation{%
Max-Planck-Institut für Plasmaphysik, D-85748 Garching, Germany
}%
\author{M. Dunne}
\affiliation{%
Max-Planck-Institut für Plasmaphysik, D-85748 Garching, Germany
}%
\author{ASDEX-Upgrade Team}
\date{\today}

\begin{abstract}
In this work, we demonstrate the utility of state representation learning applied to modeling the time evolution of electron density and temperature profiles at ASDEX-Upgrade (AUG). The proposed model is a deep neural network which learns to map the high dimensional profile observations to a lower dimensional state. The mapped states, alongside the original profile's corresponding machine parameters are used to learn a forward model to propagate the state in time. We show that this approach is able to predict AUG discharges using only a selected set of machine parameters. The state is then further conditioned to encode information about the confinement regime, which yields a simple baseline linear classifier, while still retaining the information needed to predict the evolution of profiles. We then discuss the potential use cases and limitations of state representation learning algorithms applied to fusion devices. 
\end{abstract}

\pacs{}

\maketitle 

\section{Introduction}
We investigate the use of state representation learning to model the time evolution of plasmas at ASDEX-Upgrade (AUG). As reviewed in \cite{Lesort2018StateOverview}, state representation learning (SRL) focuses on learning low dimensional features of an environment that evolve in time and are influenced by actions. An SRL model posits a system's state at a given time, $s_t$, with observations, $o_t$, which are noisy measurements of the state. The state evolves under the influence of actions, leading to future states $s_{t+1}$, which can again be measured, $o_{t+1}$.  In this work, we consider AUG to be the environment in which actions, $a_t \in \mathcal{A}$, are made at a time step $t$, where $\mathcal{A}$ is the action space. At AUG, actions are `machine control parameters', such as plasma current, magnetic field strength, gas puffing rate, etc. The change in machine parameters induces a change in the plasma state, $s_t$ to $s_{t+1}$. Full information of the true plasma state is not accessible, but diagnostic systems, such as Thomson scattering and reflectometry, provide partial and noisy observations, $o_t \in \mathcal{O}$, of the plasma state. 

\begin{figure}[h]
    \centering
    \includegraphics[scale=0.5]{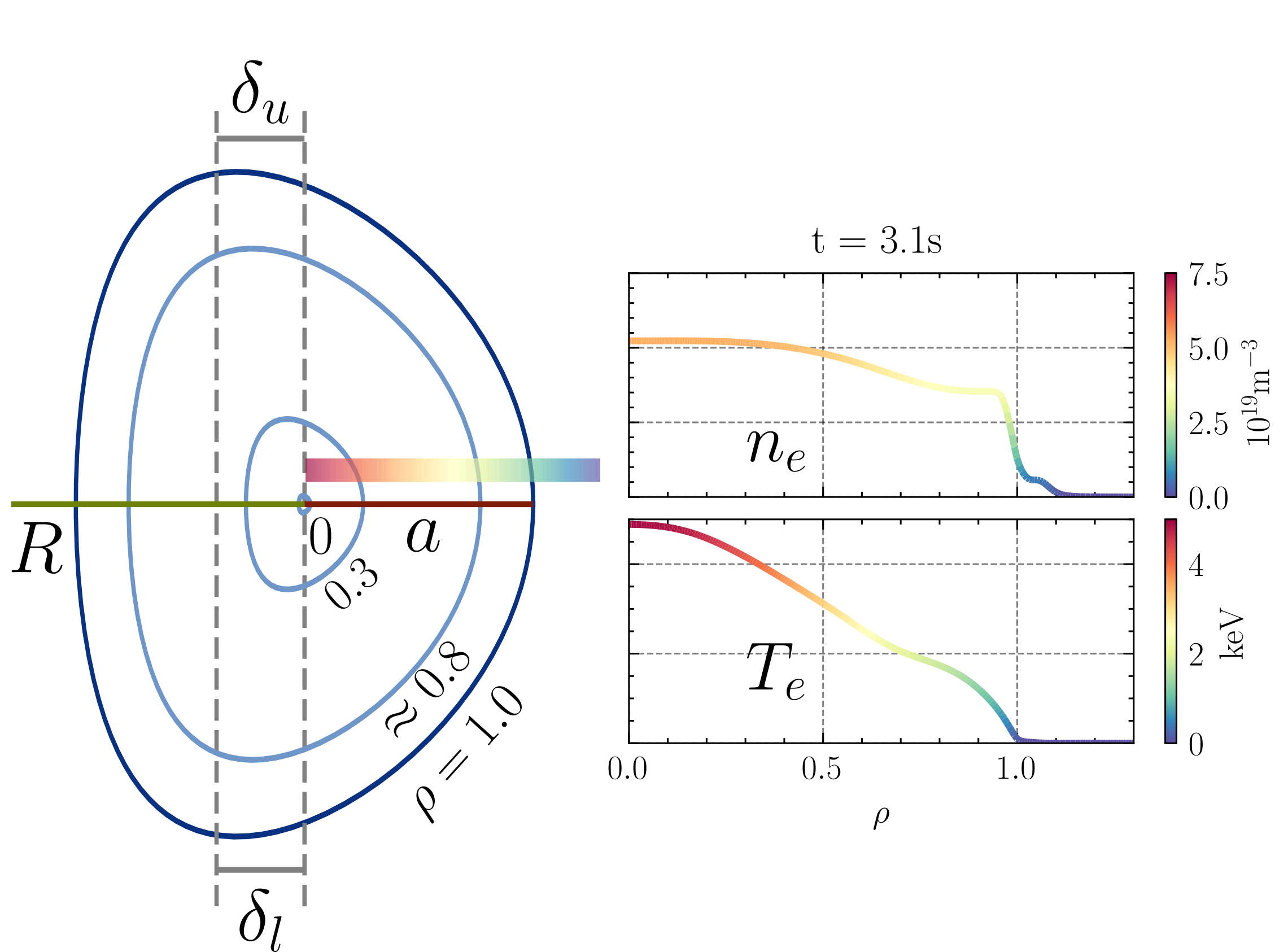}
    \caption{A representation of the plasma state at ASDEX-Upgrade. On the left, a 2D cross section of the plasma with various flux surfaces labeled by their flux surface coordinate $\rho$. The confined region of the plasma spans from the core ($\rho = 0.0$) to the separatrix ($\rho=1.0$). Machine parameters related to the shape of the plasma are labeled; the upper/lower triangularity, $\delta_{u/l}$, the major and minor radius, $R$, $a$. On the right, observations of the electron density and temperature are visualized. The main plasma kinetic profiles are typically remapped to the outer-mid-plane (location corresponding to the colorbar in the left). Flux surfaces move during the course of the discharge, as does the magnitude of the electron profiles; we seek to model the dynamics of the kinetic profiles in this work.}
    \label{fig:SRL-overview}
\end{figure}

The goal of this work is to investigate methods to learn a useful representation of the AUG plasma state, $s_t \in \mathcal{S}$, with which we can then learn a forward model, $p(s_{t+1} | s_{t}, a_{t})$, to predict the evolution and dynamics of the plasma state. In this work, a useful state representation is defined as one which represents the high dimensional observations in a state that conforms to the actual degrees of freedom of the system, i.e., a state that retains the information content of the observations.  

Compressed or lower dimensional state representations are desirable for control\cite{Hafner2023MasteringReturn, abbate_2023}, predictive inference\cite{Kit2023DevelopingJET, Zhu2022Data-drivenPrediction}, and interpretation\cite{JoyCAPTURINGVAEs}, among others. Ideally, such representations contain the essential aspects of the system. In practice, it is hard to enforce this, thus, the overarching question we seek to address is: What constitutes the objective function that learns a useful state? 

In this work, a variational autoencoder\cite{Kingma2014Auto-encodingBayes} (VAE) is used to learn a state representation of electron density and temperature profile measurements of AUG plasmas. The state representation and machine parameters are used to train a forward model to predict the dynamical evolution of the state. The end result is a model that can predict dynamical evolution of the electron density and temperature profiles directly from a sequence of machine parameters (Fig. \ref{fig:SRL-overview}). 

\section{Dataset}
The dataset used in this analysis consists of $~$1000 high confinement mode (H-mode) discharges that are non-disruptive, deuterium fuelled, and without impurities. For each pulse, observations are outer-mid-plane electron profiles of density, $n_e$ and temperature, $T_e$, and actions are machine control parameters.  


The electron profiles are obtained from the IDA\cite{Fischer2010IntegratedUpgrade} at AUG, which applies Bayesian probability theory to fit a spline to core-edge measurements originating from lithium beams, electron cyclotron emissions, Thomson scattering and interferometry. 

The following machine control parameters were selected: total plasma current, $I_\text{P}$, safety factor magnitude at 95\% flux surface , $q_{95}$, total deuterium injection rate, $D_\text{TOT}$, plasma major radius, $R$, plasma elongation $\kappa$, upper triangularity $\delta_u$, lower triangularity,  $\delta_l$, aspect ratio, $A = R/a$ (where $R$ and $a$ are the major and minor radii of the plasma), and total heating power normalized by the Martin LH-threshold scaling \cite{Takizuka2008PowerITER}, $P_\text{TOT}/P_\text{LH}$. The motivation for using the Martin scaling instead of just $P_\text{TOT}$ is that $P_\text{TOT}/P_\text{LH}$ is a normalized parameter with respect to plasma scenarios and can be applied for any device size that was used in the scaling. The chosen machine parameters are considered to be 'controllable' even though not all are strictly knobs on the tokamak, i.e., $I_\text{P}$ is not the current through the central solenoid but is a quantity that is achieved through attenuation of controllable parameters (the current through the central solenoid). The same holds for the parameters related to the plasma shape even though they are all reconstructed values of the plasma state ($\delta_l, \delta_u, A, R$ and $\kappa$).  
The machine parameters are linearly interpolated in time with respect to the IDA measurement frequency (1kHz) in order to homogenize the sampling frequencies of both actions and observations. 

Then, the combined set of observations and actions are time-wise downsampled, which transformed the time step frequency from 1kHz to 200Hz, i.e., observations and actions were selected every 5ms as opposed to the 1ms true sampling interval. This is done in part because the observations and actions chosen in this work tend to have low variability within 5ms intervals.


The data is split into training, validating and testing subsets, consisting of 853 ($\sim$5500 real-time seconds), 137, 137 discharges respectively. Additionally, discharges coming from the same shot request are binned into the same subset, which helps ensure that the training set does not include similar discharges as the validating and testing sets. The observations and actions are z-score normalized via the mean and standard deviation of the training set.  

\section{Model}
The model follows `World Modeling' as first proposed in \cite{HaWorldModels}; here an \textbf{observational} model is used to compress measurements at a given time to a latent distribution, $s_i$, and a \textbf{forward} model to evolve this distribution into the future, $s_j$ where $j>i$. Following more recent advances in World Modeling \cite{HafnerLearningPixels, Hafner2023MasteringReturn}, we additionally train the observational and forward models in one computational graph. Finally, a physics prior is introduced to guide the representation to be more physically informative.

The goal of the observational model is to learn a function that can reconstruct observations from a given state, i.e., learn $\phi(\theta_\phi): o_t \rightarrow s_t$ and $\pi(\theta_\pi): s_t \rightarrow \hat{o}_t$, where $\phi$ has parameters $\theta_\phi$. We assume that these relations are not deterministic therefore we argue to treat these function as probability distributions. To do so, we employ a VAE, a probabilistic generative model, consisting of an encoder ($\phi$) and decoder ($\pi$) distribution. The encoder and decoder distribution are parameterized by neural networks. The distributions are learned by minimizing the reconstruction error ($L^2$ norm) between observations, $o_t$, and their reconstructions, $\hat{o}_t$, in combination with a regularizing term (Kullbeck-Leibler divergence) on a prior belief of the encoder distribution, $\phi_{\theta_\phi}(s_t \mid o_t)$, resulting in our implementation of the VAE objective function: 
$$\mathcal{L}_{\textbf{obs}} =  \mathbb{E}_{\phi, \pi} \left[||\hat{o}_t - o_t ||_2 \right]+ \KL{\phi_{\theta_\phi}(s_t\mid o_t)}{p(s_t)} $$ where $p(s_t) = \mathcal{N}(0, 1)$ is our prior belief about the state distribution, and the expectation $\mathbb{E}_{\phi, \pi}$ is estimated using the \textit{re-parameterization trick} \cite{Kingma2014Auto-encodingBayes}. The prior belief about the state distribution is strictly a design choice to allow for an unconstrained model; future work would include identifying physics-based distribution as priors. The effect of different norms for the reconstruction loss was not explored in this work due to the efficiency of the $L^2$ norm. The architecture of the observational model is given in Table \ref{tab:observation-architecture}. 

\begin{table}[h]
\caption{\label{tab:observation-architecture} Observational model architecture. The observations are 1D profiles with two channels, the density and temperature. The 1D convolution and 1D transposed convolution layers are denoted as Conv. and Transp. Conv. respectively. The parameters of the convolution are denoted as (in channels=$i$, out channels=$o$, kernel width=$k$, stride=$s$), i.e., a Convolution layer with 2 in channels, 4 out channels, a kernel width of 4 and stride of 2 would be denoted as Conv. (2, 4, 4, 2). The Encoder to State has two components, denoting the mean and standard deviation of the latent variable. }
\begin{tabular}{c | c  c}
\hline
Model Component & Layer(s) & Activation \\  
\hline
Encoder & \begin{tabular}{@{}c@{}}Conv. (2, 4, 4, 2) \\ Conv. (4, 8, 4, 2) \\ Conv. (8, 16, 4, 2)\\ Conv. (16, 32, 4, 2) \end{tabular} & \begin{tabular}{@{}c@{}}ReLU \\ ReLU \\ ReLU \\ None \end{tabular} \\ 
\hline
Encoder to State & Dense (320, 8), Dense (320, 8) & None\\ 
\hline
Decoder & \begin{tabular}{@{}c@{}}Dense (8, 128) \\ Transp. Conv. (128, 16, 5, 3) \\ Transp. Conv. (16, 8, 6, 3) \\ Transp. Conv. (8, 4, 6, 3)\\ Transp. Conv. (4, 2, 6, 3) \end{tabular} & \begin{tabular}{@{}c@{}} None \\ ReLU \\ ReLU \\ ReLU \\ ReLU \end{tabular} \\ 
\hline
Output & Dense (165, 200) & None \\
\hline
\end{tabular}
\end{table}

The goal of the forward model is to predict a future state $s_{t+1}$ from the previous state $s_{t}$ and machine control parameters $a_t$, i.e., to learn the mapping $f(\theta_f): s_t, a_t \rightarrow \hat{s}_{t+1}$. Since the state $s_t$ is a distribution, the output of the forward model, $\hat{s}_{t+1}$, is the parameters of a probability distribution, also parameterized by a neural network, from which $\hat{s}_{t+1}$ is sampled. To match $s_{t+1}$ with $\hat{s}_{t+1}$ the following objective function is used:
$$\mathcal{L}_{\textbf{f}} = \mathbb{E}_{\phi, f}[\KL{\phi(s_{t+1}\mid o_{t+1})}{f(\hat{s}_{t+1}\mid s_t, a_t)}]$$
where once again, the Kullbeck-Liebler divergence is used. The architecture of the forward model is given in Table \ref{tab:forward-architecture}.

\begin{table}
\caption{\label{tab:forward-architecture} Forward Architecture. The initial layer of the forward model has size 8 (state size) + 9 (action size) $= 17$. Similar to the observational model, the forward model outputs the parameters of a distribution, therefore the last layers parameterize the mean and standard deviation.}
\begin{tabular}{c | c  c}
\hline
Model Component & Layer(s) & Activation \\ 
\hline
State to State &   \begin{tabular}{@{}c@{}}Dense(8+9, 20) \\ Dense (20, 8), Dense (20, 8) \end{tabular} &  \begin{tabular}{@{}c@{}} None \\ None \end{tabular} \\
\hline
\end{tabular}
\end{table}

\begin{figure}
    \centering
    \includegraphics[width=\columnwidth]{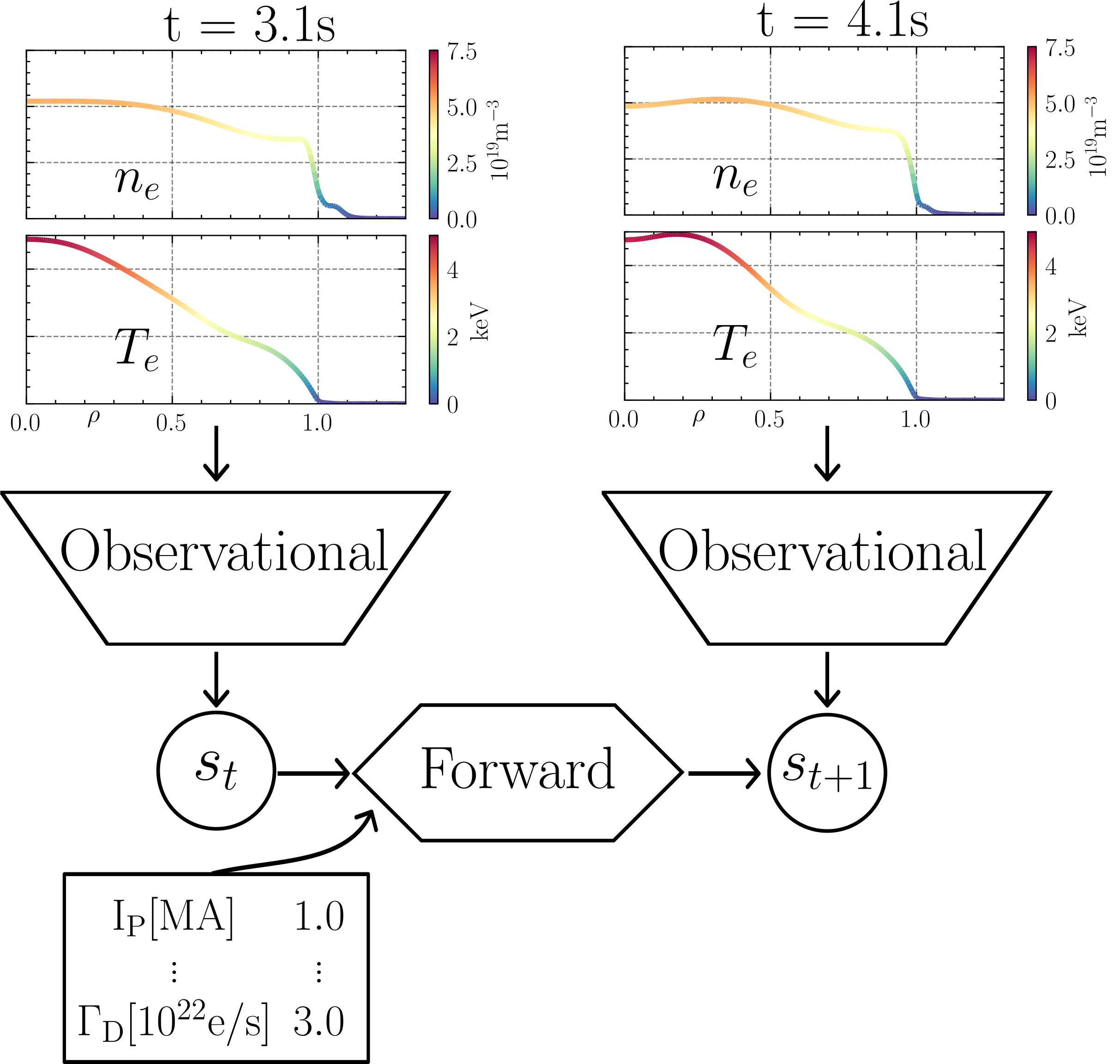}
    \caption{Graphical representation of the full model. Electron profiles are encoded via the \textit{observational} model to the state $s_t$. To predict $s_{t+1}$, the \textit{forward} model takes $s_t$ and actions $a_t$, here the plasma current $\text{I}_\text{P}$and gas puff rate $\Gamma_D$. The observational model can be used to decode the state to retrieve kinetic profiles.}
    \label{fig:enter-label}
\end{figure}

Together, a prediction of a future state, $s_{t+1}$, is made by first encoding $o_t$ to $s_t$, then transitioning from $s_t$ to $s_{t+1}$ (Fig \ref{fig:enter-label}). In this fashion, the forward model and observational model are trained simultaneously by minimizing the following objective: 
$$\mathcal{L} = \mathcal{L}_{\textbf{obs}} + \mathcal{L}_{\textbf{f}}$$
It is worth noting here that by allowing gradients to flow from $\mathcal{L}_{\textbf{f}}$ back through to the encoder of the observational model, the learned state representation is expected to retain properties that facilitate state dynamics prediction by the forward model. 

Additionally, two forms of regularization are added to the model: i) a penalty on violation of static pressure conservation in reconstruction $\propto || n_tT_t  - \hat{n_t}\hat{T_t}||_1$ and ii) the \textit{pushforward} trick from \cite{BrandstetterMESSAGESOLVERS}, where the forward model predicts $\hat{s}_{t+i}$, with $i > 1$, using the previous forward model prediction of $s_{t+i-1}$. 
Ideally, the pressure penalty encourages the observational model to encode physically consistent electron density and temperature reconstructions and was first explored in \cite{Kit2023DevelopingJET}. The pressure penalty is very similar to the reconstruction error, however we believe the pressure penalty helps to regularize the predictions of the density and temperature with respect to fluctuations around a pressure value. For example, if the reconstruction error is 0, then the pressure error is also 0. Yet, if the reconstruction error is non-zero, then the pressure error could be $\geq 0$, i.e., the fluctuations in temperature and density may even out to yield zero pressure error. If the model is wrong, we would rather the model learn to be wrong in this way.  
The \textit{pushforward} trick\cite{BrandstetterMESSAGESOLVERS} aims to stabilize auto-regressive models in long-range planning. During training, the number of time steps to rollout, $i$, is determined per mini-batch by sampling from a uniform distribution $U[0, N]$ where $N$ is the number of epochs trained thus far. The loss is only calculated between the final rollout state, $\hat{s}_{t+i}$, and corresponding state $s_{t+i}$. In other words, we cut the gradients in the unrolling stage. 

Since the space of observations $\mathcal{O}$ is constrained to $\mathcal{R}^+$, the output of the observation model is clamped to output only positive real values during training. This is done by clamping the output of the observational decoder. 

Due to the competing objectives and to normalize the reconstructed pressure penalty with respect to the remaining objectives, we found it useful to weight individual components with scaling factors. Training hyperparameters and objective penalty weights are given in Table \ref{tab:hyperparameters}. A rigorous search for optimal hyperparameters, including state size, was not conducted. However, the final configurations \ref{tab:hyperparameters} were selected among others by obtaining the lowest error on the validation dataset.

\begin{table}[h]
    \caption{\label{tab:hyperparameters} Objective weights and training parameters used for the SRL model. All weights are scalar  multiplied by their corresponding objective value per mini-batch update. $\text{KL}_{\textbf{obs}}$ is applied to the KL term in $\mathcal{L}_\textbf{obs}$. $\text{KL}_{\textbf{f}}$ is applied to the KL term of in $\mathcal{L}_{\textbf{f}}$. $L^1_{o_t}$ is applied to the $L^1$ term of $\mathcal{L}_\textbf{obs}$. $L^1_{p_t}$ is applied to the $L^1$ pressure penalty.}
    \begin{tabular}{m{1cm}|m{1.75cm} |m{1.75cm} |m{1.5cm} | m{1.5cm}}
        \hline
         \multicolumn{5}{c}{Objective Weights} \\
         \hline
         Name & $\text{KL}_{\textbf{obs}}$ & $\text{KL}_{\textbf{f}}$ & $L^1_{o_t}$ & $L^1_{p_t}$\\ 
         Value &  0.01 & 1.0 & 100 & 0.0001\\
         \hline
         \multicolumn{5}{c}{Training Hyperparameters} \\
         \hline
         Name & Batch Size & Optimizer & \multicolumn{2}{| c }{Learning Rate} \\
         Value & 8 & Adam \cite{Kingma2014Adam:Optimization} & \multicolumn{2}{| c }{0.02}\\ 
         \hline
    \end{tabular}
    
\end{table}

\section{Results}
\subsection{Observational Model}
\begin{figure}
    \centering
    \includegraphics[width=\columnwidth]{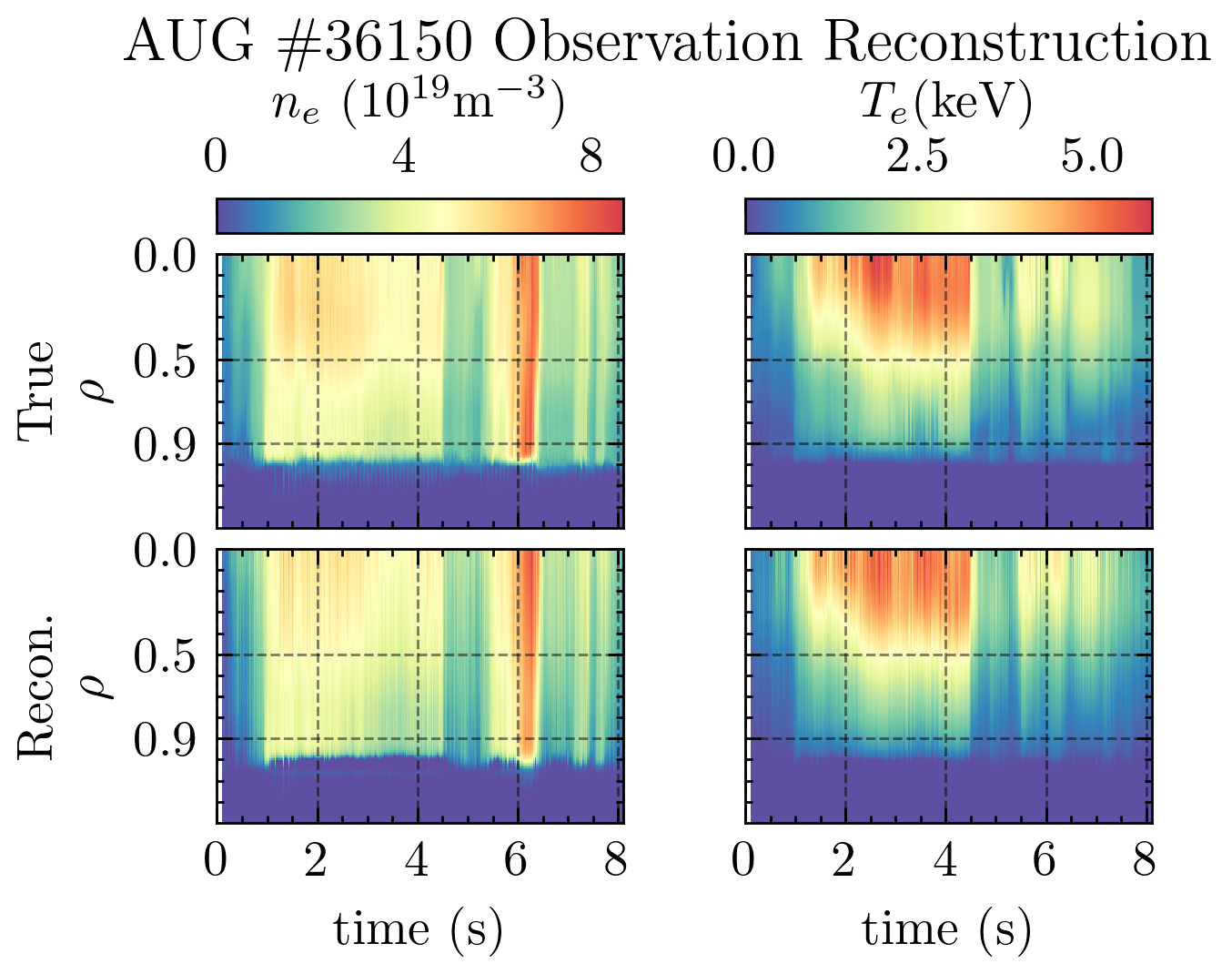}
    \caption{Observational model's reconstruction of AUG discharge \#36150. The left and right figures show the density and temperature profile evolutions, respectively. The top and bottom plots show the true profiles and model reconstruction, respectively. The x and y-axis on all figures are the same. The reconstruction error is similar to that of the average of the test set discharges, as the MAE of the density and temperature profiles averaged over this discharge are $0.32 \pm 0.29$ ($10^{19}$m$^{-3}$) and $0.12 \pm 0.12$ (keV), respectively. Mean and standard deviation of the errors are calculated over 100 sample reconstructions. Respective errors at $\rho = 0.0, 0.5, 0.9, 1.0$ are given in Table \ref{tab:obs_mae_rhos}.}
    \label{fig:obserational-reconstruction-36150}
\end{figure}

The quality of the observational model can be determined by comparing the observation with the reconstruction (Figure \ref{fig:obserational-reconstruction-36150}). For the test-set discharges, a mean absolute error (MAE) between reconstruction and observation has a mean of $0.28 \pm 0.13$ $10^{19}$m$^{-3}$  and $0.11 \pm 0.07$ (keV) for density and temperature, respectively. These lossy compression results are expected, as the state space is only 8-D while the observational space is 400 points for each time step. Increasing the state space dimensionality would likely yield lower reconstruction error. The average reconstruction error (MAE) of the test set discharges for $\rho=0.0, 0.5, 0.9, 1.0$ is given in Table \ref{tab:obs_mae_rhos}. 

We find the reconstruction quality of the observational model to be sufficiently accurate to proceed with the forward model. 

\begin{table}
\caption{\label{tab:obs_mae_rhos} The MAE of the observational model's reconstructions for various $\rho$. The MAE for AUG \#36150 (Fig. \ref{fig:obserational-reconstruction-36150}) is provided to compare with the average over the test set discharges. Standard deviations for all values are calculated over 100 sample reconstructions, given the injection of noise in the VAE.}
\begin{tabular}{c | c | c |  c | c }
\hline
\multicolumn{5}{c}{Average MAE over test set discharges} \\
\hline
$\rho$ & 0.0 & 0.5 & 0.9 & 1.0 \\ 
$n_e$ ($10^{19}$m$^{-3}$) & 0.46 $\pm$ 0.6 & 0.31 $\pm$ 0.13 & 0.33 $\pm 0.17$ & $0.41 \pm 0.28$\\ 
$T_e$ (keV) & 0.3 $\pm$ 0.12 & 0.11 $\pm$ 0.09 & 0.06 $\pm$ 0.03 & 0.03 $\pm$ 0.06 \\
\hline 
\multicolumn{5}{c}{AUG \#36150 (Fig. \ref{fig:obserational-reconstruction-36150})} \\
\hline 
$n_e$ ($10^{19}$m$^{-3}$) & $0.23 \pm 0.18$ &  $0.43 \pm 0.27$ &  $0.45 \pm 0.27$ &  $0.78 \pm 0.6$ \\ 
$T_e$ (keV) & $0.26 \pm 0.17$ &  $0.16 \pm 0.12$ &  $0.1 \pm 0.06$ &  $0.06 \pm 0.06$ \\
\hline
\end{tabular}
\end{table}

\subsection{Forward Model}
\begin{figure}[h]
    \centering
    \includegraphics[width=\columnwidth]{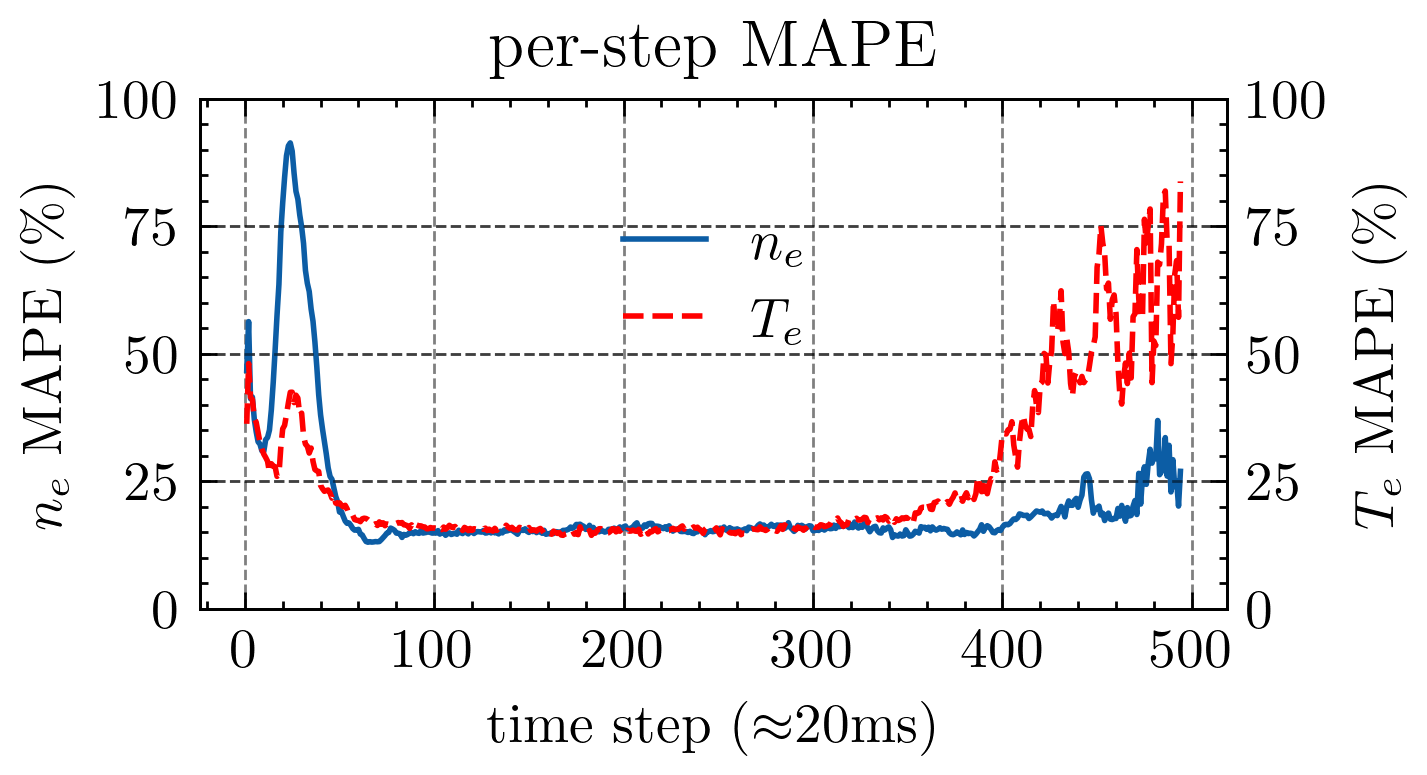}
    \caption{The test set forward model error (MAPE) as a function of time. The error per step is calculated as the average over the density and temperature profiles up to $\rho\leq 1.0$ for that step. The reason for radial cutoff is the very low values of temperature and density at $\rho > 1.0$. The spikes at the beginning and end are likely due the discharge entering and exiting H-mode, where the density and temperature rapidly change and are therefore difficult to precisely match. }
    \label{fig:cumulation-error-forward}
\end{figure}

To determine the predictive quality of the forward model on AUG discharges, we first encode the observations at $t=0$ to a state $s_0$ via the observational model, then the forward model rolls $s_0$ out to the final time step using true actions and its own predictions of $\hat{s}_{t>0}$. Each $\hat{s}_t$ is then decoded into profiles via the observational decoder. The MAE of all time steps over all discharges for the forward model is  $0.97 \pm 0.55$ $10^{19}$m$^{-3}$  and $0.35 \pm 0.24$ (keV) for density and temperature, respectively. The average reconstruction error (MAE) of the test set discharges for $\rho=0.0, 0.5, 0.9, 1.0$ is given in Table \ref{tab:trans_mae_rhos}. The average percentage reconstruction error (MAPE) is given in Table \ref{tab:trans_mape_rhos}. Both the MAE and MAPE are reported due to large variations radially of the magnitude of density and temperature, i.e., at $\rho>1.0$ density and temperature are relatively low compared to $\rho=0.0$. The mean accumulation of error does not rapidly increase over time for test-set discharges (Fig. \ref{fig:cumulation-error-forward}).

\begin{table*}
\caption{\label{tab:trans_mae_rhos} The MAE of the forward model's reconstructions for various $\rho$. The MAE of the AUG discharges visualized in this work are provided for comparison with the average over the test set discharges. Standard deviations for all values are calculated over 100 sample reconstructions.}
\begin{tabular}{c | c | c |  c | c }
\hline
\multicolumn{5}{c}{Average MAE over test set discharges} \\
\hline
& $\rho = 0.0$ & 0.5 & 0.9 & 1.0 \\ 
$n_e$ ($10^{19}$m$^{-3}$) & 1.51 $\pm$ 0.83 & 1.13 $\pm$ 0.76 & 1.07 $\pm 0.61$ & $0.68 \pm 0.43$\\ 
$T_e$ (keV) & 0.96 $\pm$ 0.72 & 0.36 $\pm$ 0.24 & 0.15 $\pm$ 0.11 & 0.05 $\pm$ 0.09 \\
\hline 
\multicolumn{5}{c}{AUG \#34828 (Fig. \ref{fig:forward-reconstruction-34828})} \\  
$n_e$  & $0.597 \pm0.57$ &  $0.39 \pm 0.36$ &  $0.38 \pm 0.28$ &  $0.25 \pm 0.21$ \\ 
$T_e$ & $0.87 \pm 0.33$ &  $0.11 \pm 0.1$ &  $0.07 \pm 0.05$ &  $0.04 \pm 0.02$ \\
\hline 
\multicolumn{5}{c}{AUG \#36022 (Fig. \ref{fig:forward-reconstruction-36022})} \\
$n_e$  & $0.67 \pm 0.57$ &  $0.83 \pm 0.26$ &  $0.74 \pm 0.31$ &  $0.5 \pm 0.37$ \\ 
$T_e$  & $0.54 \pm 0.41$ &  $0.17 \pm 0.1$ &  $0.09 \pm 0.06$ &  $0.02 \pm 0.02$ \\
\hline 
\multicolumn{5}{c}{AUG \#36150 (Fig. \ref{fig:model-limitations-36150})} \\
$n_e$ & $2.57 \pm 1.02$ &  $1.86 \pm 0.93$ &  $1.76 \pm 0.79$ &  $2.99 \pm 0.84$ \\ 
$T_e$  & $1.22 \pm 1.05$ &  $0.55 \pm 0.39$ &  $0.32 \pm 0.28$ &  $0.54 \pm 0.21$ \\
\hline 
\multicolumn{5}{c}{AUG \#36669 (Fig. \ref{fig:model-limitations-36669})} \\
$n_e$  & $1.21 \pm 1.23$ &  $1.01 \pm 0.86$ &  $0.88 \pm 0.79$ &  $0.74 \pm 0.5$ \\ 
$T_e$  & $0.87 \pm 0.51$ &  $0.40 \pm 0.45$ &  $0.15 \pm 0.09$ &  $0.1 \pm 0.02$ \\
\hline
\end{tabular}
\end{table*}

\begin{table*}
    \caption{\label{tab:trans_mape_rhos} The mean-absolute percentage error (MAPE) of the forward model's reconstructions for various $\rho$. The MAPE is calculated as the $L^1$ difference between the predicted and true value, divided by the true value. The MAPE of the AUG discharges visualized in this work are provided for comparison with the average over the test set discharges. Standard deviations for all values are calculated over 100 sample reconstructions. Large deviations (MAPE $> 100\%$) in the edge are expected, as the temperature and density tend to be relatively low ($n_e < 10^{18}\text{m}^{-3}, T_e < 140$eV).}
    \begin{tabular}{c | c | c |  c | c }
    \hline
    \multicolumn{5}{c}{Average MAPE over test set discharges} \\
    \hline
    $\rho$ & 0.0 & 0.5 & 0.9 & 1.0  \\
    $n_e$ (\%) & 28.89 $\pm$ 20.37 & 28.55 $\pm$ 19.97 & 34.31 $\pm$ 29.76 & 376.6 $\pm$ 326.6 \\
    $T_e$ (\%) & 65.82 $\pm$ 101.6 & 26.53 $\pm$ 17.71 & 73.60 $\pm$ 167.2 & 208.7 $\pm$ 353.4 \\
    \hline
    \multicolumn{5}{c}{AUG \#34828 (Fig. \ref{fig:forward-reconstruction-34828})} \\
    $n_e$ & 10.73 $\pm$ 14.63 & 12.47 $\pm$ 15.57 & 12.05 $\pm$ 10.23 & 17.73 $\pm$ 15.60 \\
    $T_e$ & 31.33 $\pm$ 15.38 & 13.67 $\pm$ 15.33 & 24.29 $\pm$ 27.46 & 52.74 $\pm$ 79.44 \\
    \hline
    \multicolumn{5}{c}{AUG \#36022 (Fig. \ref{fig:forward-reconstruction-36022})} \\
    $n_e$ & 12.97 $\pm$ 10.18 & 15.775 $\pm$ 10.15 & 15.837 $\pm$ 15.82 & 22.33 $\pm$ 54.52 \\
    $T_e$ & 21.70 $\pm$ 16.77 & 13.788 $\pm$ 10.32 & 31.516 $\pm$ 97.79 & 43.27 $\pm$ 129.8 \\
    \hline
    \hline
    \multicolumn{5}{c}{AUG \#36150 (Fig. \ref{fig:model-limitations-36150})} \\
    $n_e$ & 53.9 $\pm$ 53.54 & 44.58 $\pm$ 47.45 & 54.44 $\pm$ 56.12 & 263.6 $\pm$ 147.1 \\
    $T_e$ & 42.76 $\pm$ 46.58 & 27.17 $\pm$ 21.95 & 51.54 $\pm$ 57.96 & 521.5 $\pm$ 177.8 \\
    \hline
    \hline
    \multicolumn{5}{c}{AUG \#36669  (Fig. \ref{fig:model-limitations-36669})} \\
    $n_e$ & 33.14 $\pm$ 46.30 & 32.66 $\pm$ 46.56 & 40.44 $\pm$ 84.46 & 104.4 $\pm$ 311.2 \\
    $T_e$ & 25.77 $\pm$ 17.39 & 17.74 $\pm$ 16.20 & 18.59 $\pm$ 20.06 & 22.31 $\pm$ 20.98 \\
    \hline
    
\end{tabular}
\end{table*}

\begin{figure}
    \centering
    \includegraphics[width=\columnwidth]{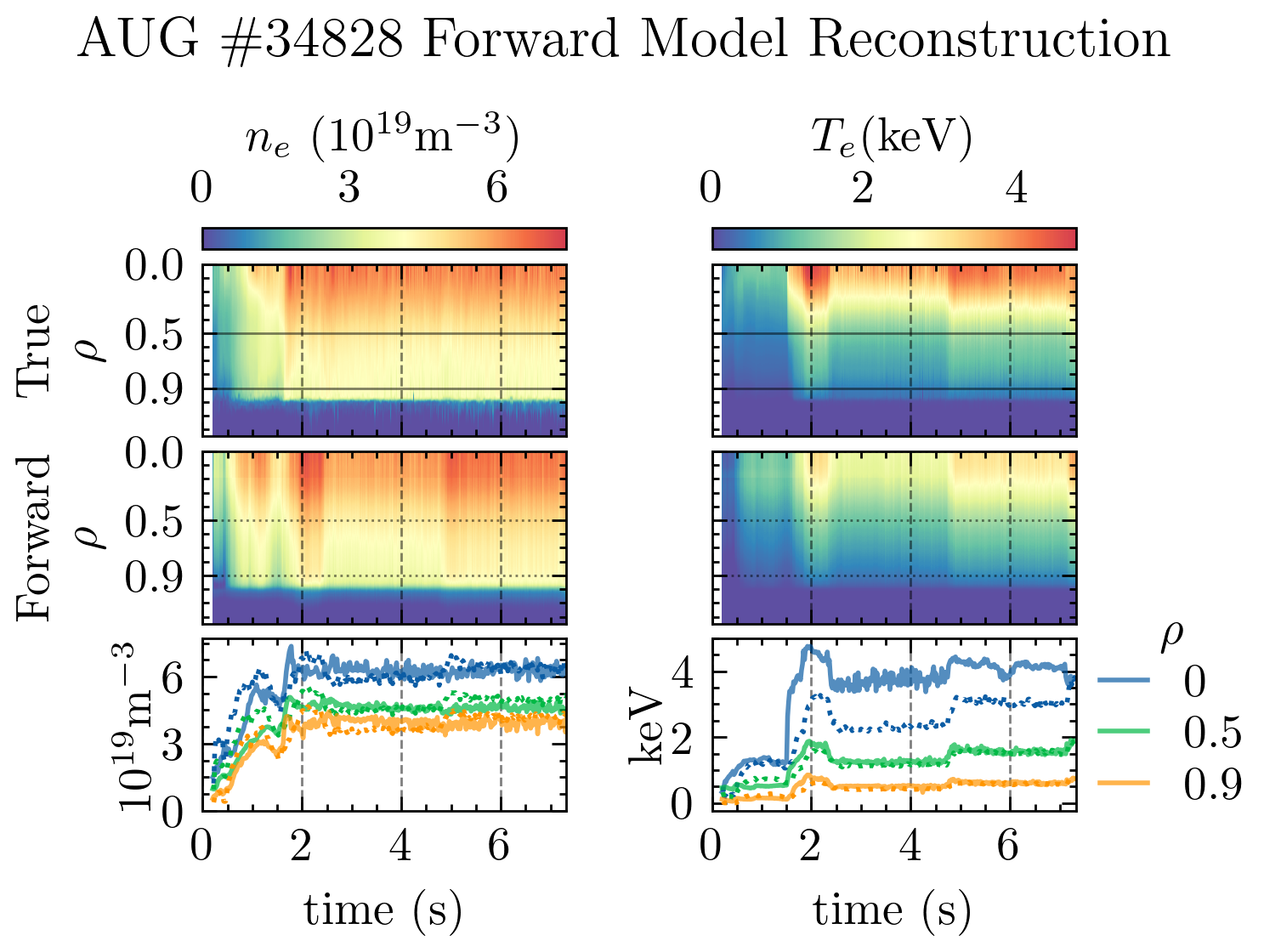}
    \caption{AUG \#34828 comparison of true (top) and forward model predicted (middle) electron density  and temperature profiles are plotted along with time traces of the radial values at $\rho=0.0, 0.5$ and 0.9 (bottom). All rows in the left column are associated with density, the right with temperature. The solid/dotted grid lines in the top/middle plot correspond to the solid/dotted traces in the bottom plot. The initial state $s_0$ is sampled from the encoder of the observational model given the profiles at $t_0$. The initial state is then propagated in time alongside actions and previous state prediction via the forward model. The density prediction is worse at the beginning of the pulse, likely due to the fluctuation feedback of the gas flow rate, but eventually stabilizes to the true value. The resulting reconstructions of the predicted states from the forward model demonstrate the capability to handle sufficiently complex plasma scenarios. Respective errors of the density and temperature at $\rho = 0.0, 0.5, 0.9, 1.0$ are given in Table \ref{tab:trans_mae_rhos} \& \ref{tab:trans_mape_rhos}.}
    \label{fig:forward-reconstruction-34828}
\end{figure}

\begin{figure}
    \centering
    \includegraphics[width=\columnwidth]{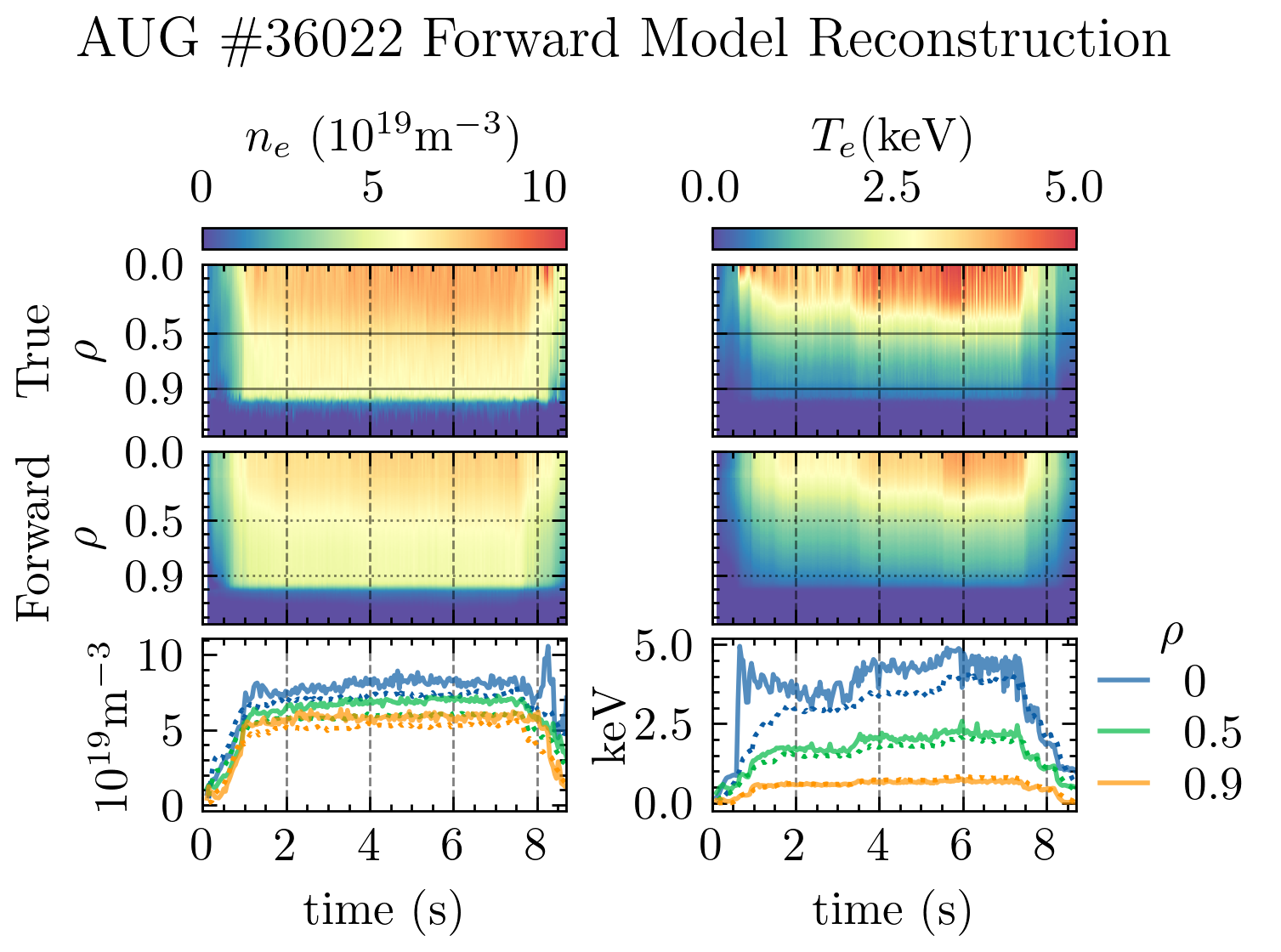}
    \caption{AUG \#36022 comparison of true (top) and forward model predicted (middle) electron density  and temperature profiles are plotted along with time traces of the radial values at $\rho=0.0, 0.5$ and 0.9 (bottom). All rows in the left column are associated with density, the right with temperature. The solid/dotted grid lines in the top/middle plot correspond to the solid/dotted traces in the bottom plot.  The initial state $s_0$ is sampled from the encoder of the observational model given the profiles at $t_0$. The initial state is and then propagated in time alongside actions and previous state prediction via the forward model. In this discharge, the total power input is increased step-wise, leading to a step-wise increase in core electron temperature, which match reconstructions of the predicted state.Respective errors of the density and temperature at $\rho = 0.0, 0.5, 0.9, 1.0$ are given in Table \ref{tab:trans_mae_rhos} \& \ref{tab:trans_mape_rhos}.}
    \label{fig:forward-reconstruction-36022}
\end{figure}

The forward model is able to capture the profile evolution for various discharges, for example, a plasma scenario with feedback density control (Fig. \ref{fig:forward-reconstruction-34828}), as well as a power step-wise ramp up (Fig. \ref{fig:forward-reconstruction-36022}). 


\subsection{Forward Model with Auxiliary Regressor}
Expanding on previous work \cite{Kit2023DevelopingJET}, we further regularize the state representation by learning a mapping $r(\theta_r): s_t \rightarrow P_\text{TOT}/P_\text{LH}, \tau_E$, where $P_\text{TOT}/P_\text{LH}$ is the aforementioned normalized power action, and $\tau_E$ the global confinement time. Inspired by \cite{JoyCAPTURINGVAEs}, we split $s_t$ into two sub-spaces $s_{c, t}$ and $s_{\textbf{/}c, t}$, and apply the mapping only on $s_{c, t}$. The dimensionality of $s_{c, t}$ is 2, one dimension for each regressed variable, and the dimensionality of $s_{\textbf{/}c, t}$ is kept to 6 so that the total dimensionality of $s_t$ is preserved from the previous experiment. Then, the mapping $r(\theta_r)$ is made to be linear with respect to the regressed variables $P_\text{TOT}/P_\text{LH}$ and $\tau_E$, i.e., a diagonal matrix that maps one dimension of $s_{c,t}$ to $P_\text{TOT}/P_\text{LH}$ and the other remaining dimension to $\tau_E$. The original loss function $\mathcal{L}$ then gains the following additional $\mathcal{L}$1 loss term: $$\mathbb{E}_{r, \phi}[||a_{c, t} - \hat{a}_{c,t} ||_1]$$
where $a_{c,t}$ and $\hat{a}_{c,t}$ are the true regressed variables and their reconstructions, respectively. 

\begin{figure}
    \centering
    \includegraphics[width=\columnwidth]{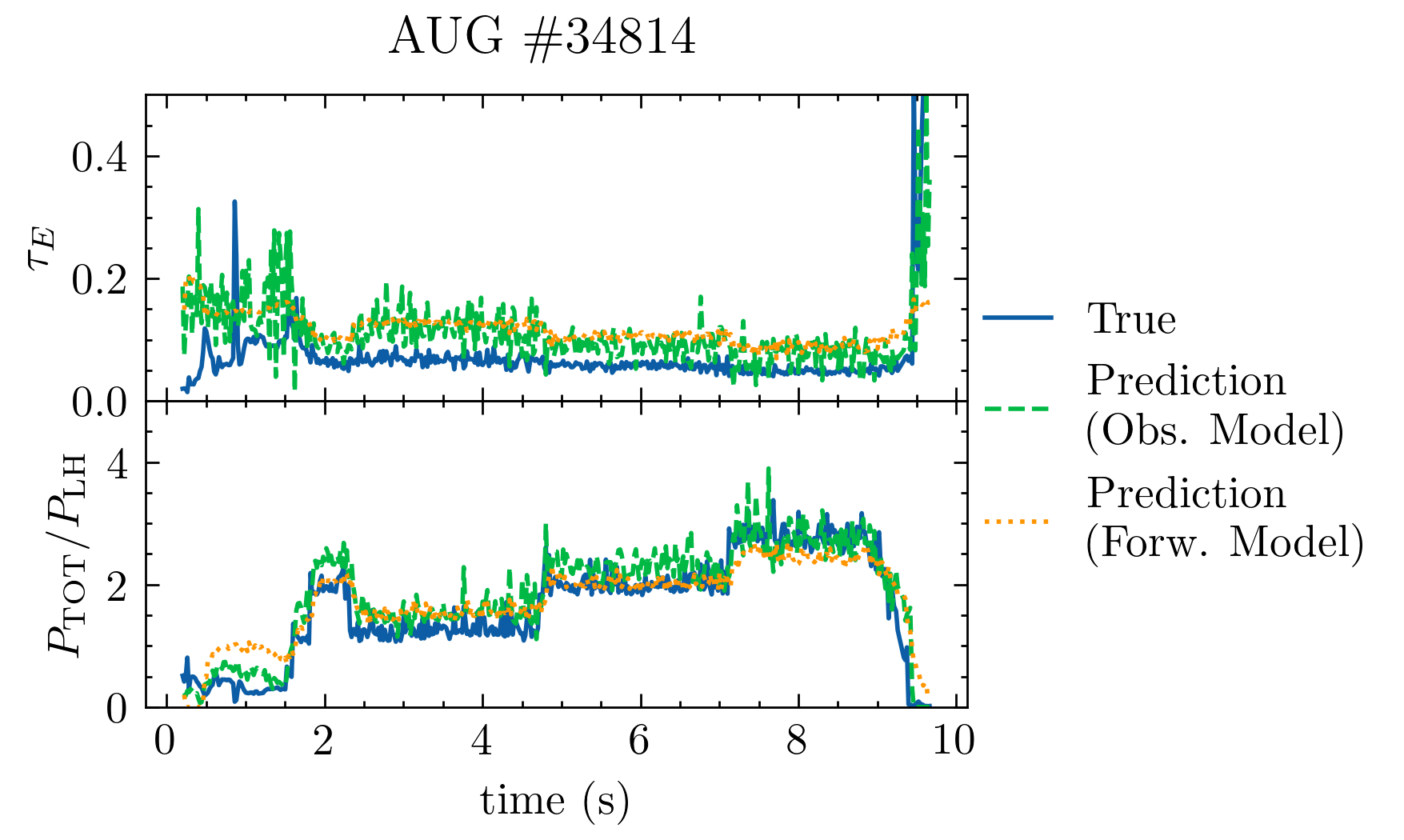}
    \caption{The predicted and true time traces of $\tau_E$ (top) and $P_\text{TOT}/P_\text{LH}$ (bottom) from AUG \#34814. The predictions of the observational model (Obs.) are obtained by encoding observations to $s_t$ and applying the auxiliary mapping. The predictions of the forward model (Forw.) are obtained by encoding the first observation to an initial state, i.e., $o_0 \rightarrow s_0$, then rolling out with the forward model until the last action. }
    \label{fig:aux-reg-recon}
\end{figure}

With the additional regressor, it was seen that the model can infer the confinement time and power variable via an observed a state $s_t$. The observed state can be encoded by either the observational model or predicted as before using the forward model (Fig. \ref{fig:aux-reg-recon}). 

While the reconstructed value of $\tau_E$ tends to be higher than the true value, the reconstructed $P_\text{TOT}/P_\text{LH}$ values are quite close to true observations. The elevated $\tau_E$ predictions might be caused by biases originating from the beginning and end of the plasma discharges and will be investigated in futures studies. The main message in this proof-of-principle work is to demonstrate the attachment of semantically meaningful information from the plasma state to the trained state representation with the auxiliary regression modules. 

\begin{figure}[h]
    \centering
    \includegraphics[width=\columnwidth]{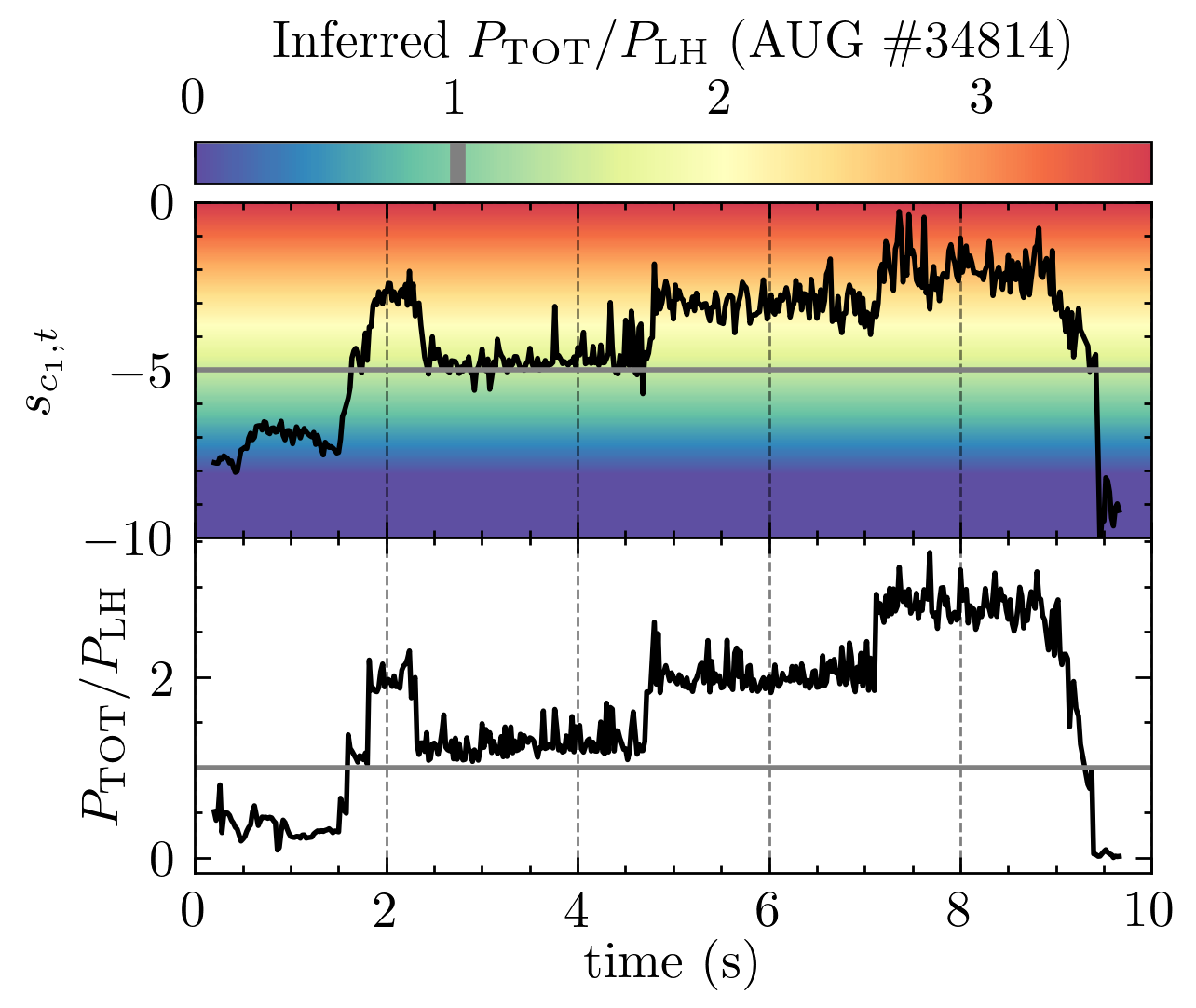}
    \caption{\textbf{Top}: The time trace of state dimension $s_{c_1, t}$ is encoded by the forward model using the actions of AUG \# 34814. The horizontal line (colorbar vertical) marks the value of $s_{c_1, t}$ which corresponds to an inferred $P_\text{TOT}/P_\text{LH} = 1$. The coloring is found by applying the auxiliary regressor to the range of $s_{c_1, t} \in [-10, 0]$. Due to the linear capacity of the auxiliary mapping, the output of the mapping on $s_{c_1, t}$ on the interval $[-10, 0]$ does not change in time, nor does it change with respect to any other state variable. Like \cite{2021PlasmaAttention}, we arrive at a model that can predict different regimes, albeit in very different fashion. \textbf{Bottom}: The time trace of  $P_\text{TOT}/P_\text{LH}$ for AUG \#34814, with horizontal line marking where $P_\text{TOT}/P_\text{LH} = 1$.} 
    \label{fig:aux-sweep}
\end{figure}

Since one dimension of $s_{c,t}$ is linear with respect to $P_\text{TOT}/P_\text{LH}$, we receive a simplified H-mode classifier without additional training. Assuming that $P_\text{TOT}/P_\text{LH}$ is sufficiently accurate in quantifying the presence of H-mode, i.e, a plasma is in H-mode when $P_\text{TOT}/P_\text{LH} \geq 1$ \textit{and} sufficient predictive quality of the auxiliary regressor, then via the linear mapping $r$, there exists an equivalent threshold, $s_{c, t} > H_\text{thresh}$ within the relevant power dimension of $s_{c, t}$, (Fig. \ref{fig:aux-sweep}). 

\begin{figure}
    \centering
    \includegraphics[width=\columnwidth]{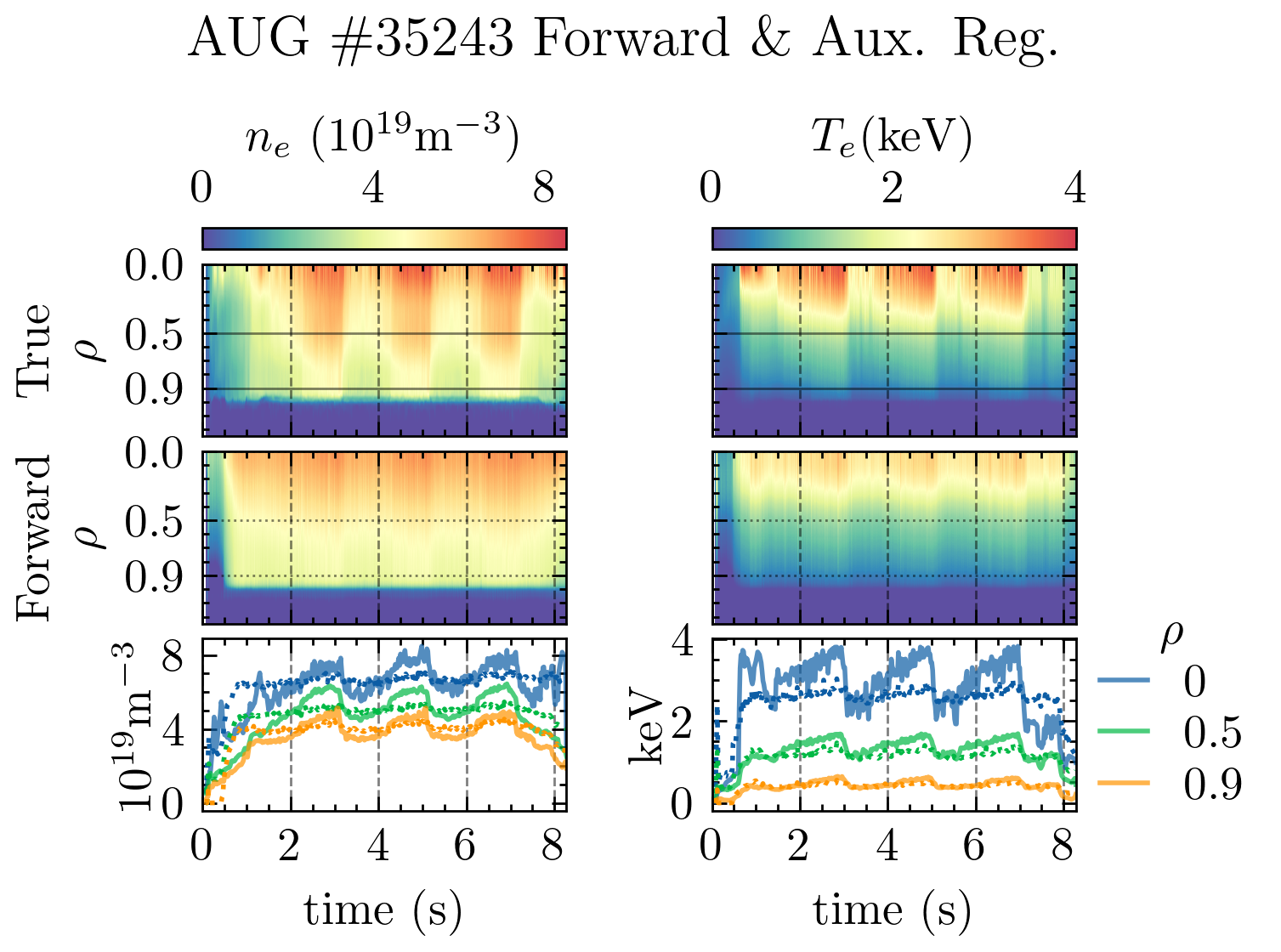}
    \caption{Even with the auxiliary regressor, the state representation and forward model are still able to capture complex time evolution of AUG discharges. The true density and temperature time traces at various $\rho$ are opaque to show the slight differences with predicted time traces. }
    \label{fig:forward-reconstruction-aux-reg-35243}
\end{figure}

However, the additional objective comes at cost, as we observed small penalties on the reconstructive quality in the forward and observational models (Fig. \ref{fig:forward-reconstruction-aux-reg-35243}). This is likely due to the two dimensions no longer free to compress information \textit{only} pertaining to profile reconstruction information. It is likely that increasing the size of the state would resolve this.

\subsection{Model Limitations}

\begin{figure}
    \centering
    \includegraphics[width=\columnwidth]{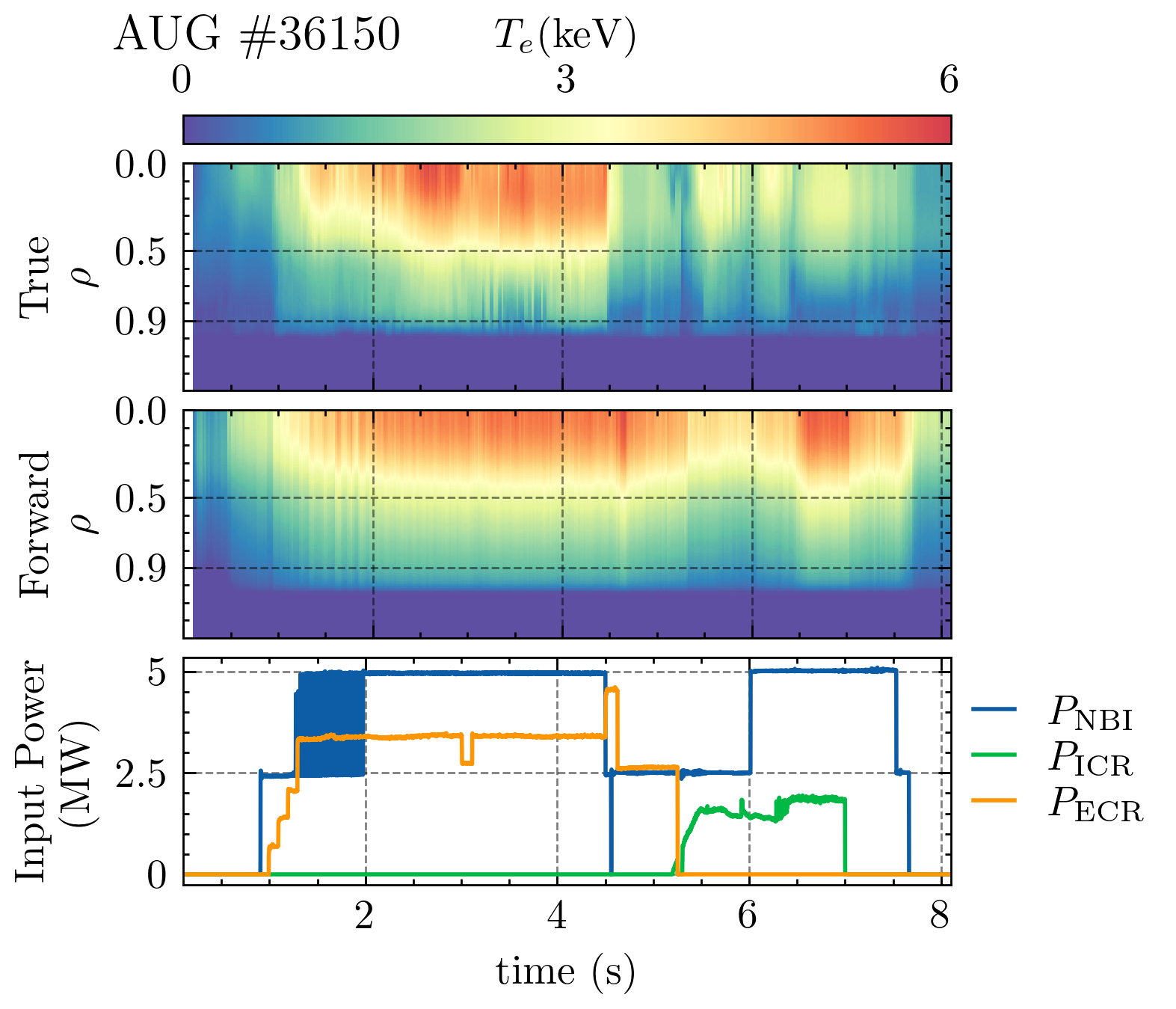}
    \caption{In AUG \#36150, the power starts as mainly NBI and ECR driven, however at 4.5 seconds, NBI is rapidly cut off and supplemented with ICR. Since $P_\text{TOT}/P_\text{LH}$ is the only available power variable to the forward model, it observes a steady stream of power, which does not induce major changes in the inferred plasma state. It is likely that including separate variables for each power parameter would increase the resilience of the model to similar discharges. Respective errors of the density and temperature at $\rho = 0.0, 0.5, 0.9, 1.0$ are given in Table \ref{tab:trans_mae_rhos}. Predictions obtained from the model without the auxiliary regressor.}
    \label{fig:model-limitations-36150}
\end{figure}

\begin{figure}
    \centering
    \includegraphics[width=\columnwidth]{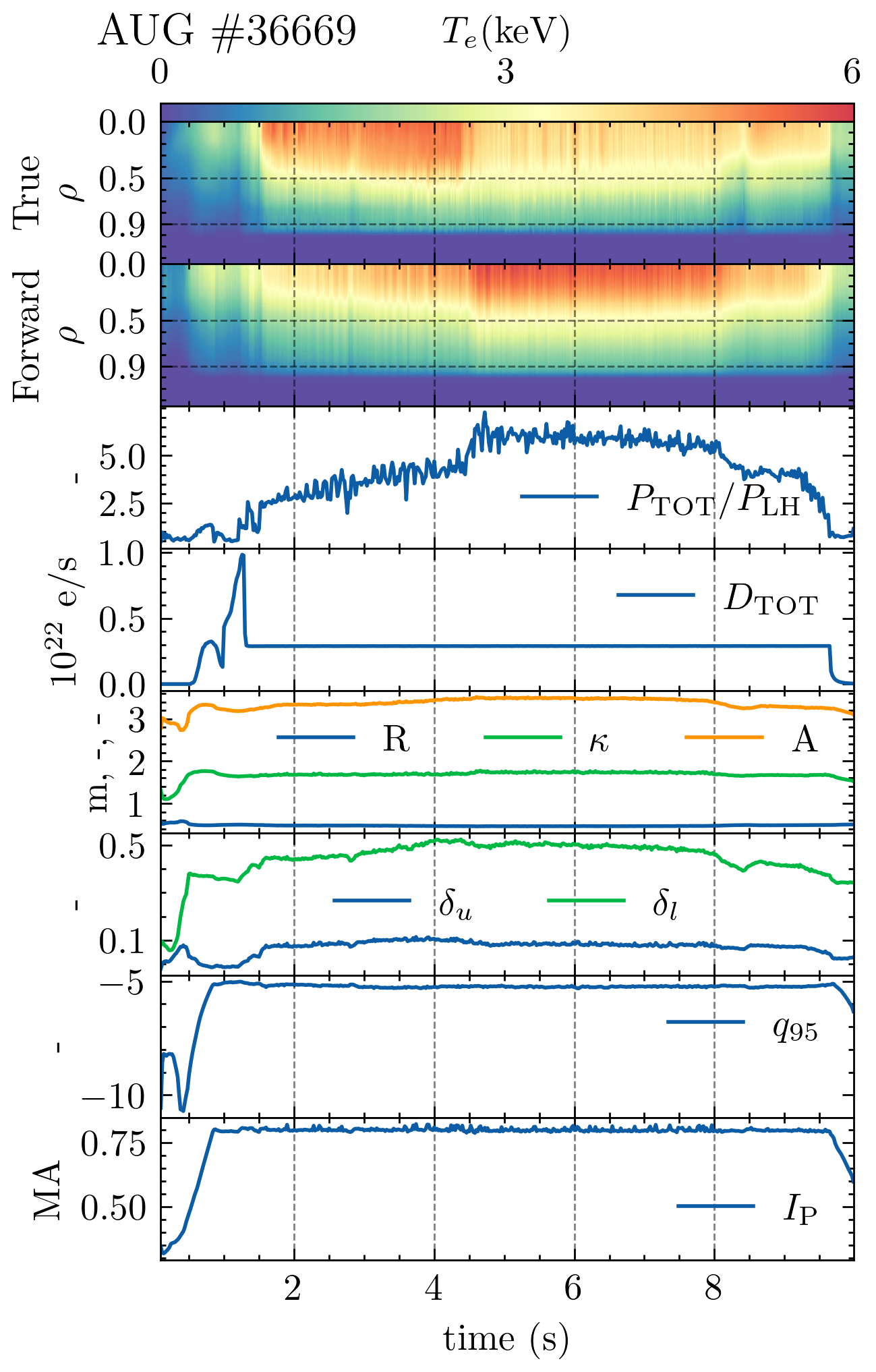}
    \caption{AUG \#36669, the top two plots are the true and forward model predicted temperature profiles and remaining plots are the corresponding actions for the pulse. The y-axis for the actions figures are the units, and their corresponding name is given within the plot. In AUG \#36669, the device is configured to test actively cooled divertor plates, and accumulation of impurities lead to an increase in core radiated power\cite{Neu2013OverviewUpgrade}, dropping the core temperature. The actions supplied to the model do not sufficiently encode this information and the model does not predict the temperature decrease. It is likely that including additional actions that are more correlated with detached/attached plasmas would increase the resilience of the model, such as those used in \cite{Zhu2022Data-drivenPrediction}. Respective errors of the density and temperature at $\rho = 0.0, 0.5, 0.9, 1.0$ are given in Table \ref{tab:trans_mae_rhos}. Predictions obtained from the model without the auxiliary regressor.}
    \label{fig:model-limitations-36669}
\end{figure}

The forward model is limited by the type and distribution of machine parameters that is provided. For example, as only the total heating power is provided, any discharges where input power mixtures is varied midshot are subject to relatively large prediction errors (Fig. \ref{fig:model-limitations-36150}). Another example is strong tungsten accumulation in the plasma, such as observed in some of the actively cooled divertor experiments (Fig. \ref{fig:model-limitations-36669}). Since the actions selected in this work do not show the corresponding variations seen on these discharges, the forward model will mispredict the resulting profiles. 


\section{Discussion}

In this work, we have demonstrated the utility of state representation learning towards learning the plasma state at ASDEX-Upgrade. Our proposed model is able to predict the electron density and temperature profiles from machine parameters only. Additionally, we demonstrate the functionality of learning a state representation by incorporating a simplified H-mode classifier into the model while retaining the ability to predict the time evolution of plasma profiles. 

The forward model developed in this work has a very limited capacity (Table \ref{tab:forward-architecture}). It might be useful to improve the forward model's capacity, e.g., via a recurrent neural network as demonstrated in \cite{Hafner2023MasteringReturn}. An alternative approach was demonstrated in \cite{KarlDEEPDATA, Becker-EhmckSwitchingFiltering}, where a deep learning variation on Kalman filters is learned by sampling the \textit{transition} between states with a VAE. Also, our approach to using a linear transform for time-stepping latent representations has some commonalities with the work in \cite{Lusch2018DeepDynamics}, where Koopman operator theory is used to guide auto-encoders into learning Koopman eigenfunctions from data, i.e., the latent space has globally linear dynamics. A key difference being that we predict the mean and standard deviation of a distribution, which we sample at each time step. Here, perhaps, there is a connection to latent neural stochastic differential equation models\cite{RyzhikovLatentDetection} (NSDE), where our autoregressive formulation can be considered a crude discretization of such an NSDE.


The cumulative error plot (Fig. \ref{fig:cumulation-error-forward}) is in some sense troubling. We expected to see a step wise accumulation of error, as with traditional forward model predictors. Interestingly, it appears that the observational model encodes information about the machine parameters, even though this information is only propagated through the forward model. As a result, when the plasma reaches flat top, the forward model likely pulls the state prediction to a previously seen steady state representation. We believe this has to do with \textit{verifiability} \cite{balsells-rodas_identifiability_2023}, and appears to be both a feature and a bug. An open question is then to what extent the latent dynamic model simply leans to propagate the state through a series of steady states rather than (ideally) predicting the temporal evolution of the plasma. Along this line, we hypothesise that one could learn a steady state model within a DIVA/CCVAE-like framework \cite{ilse2019diva, JoyCAPTURINGVAEs}, as in \cite{Kit2023DevelopingJET}, treat all time slices as steady state, and ultimately forgo the forward model. Future studies will investigate steady-state and dynamical models within the context of verifiability.

We believe an important question is what information, and at what frequency, is needed to predict future plasma states. Additional questions arise; assuming the true plasma state is Markovian, as we suspect, then what observables are necessary to capture that? Also, if we can approximate the true state sufficiently well in a low dimensional representation, then a) what observations are used to learn such a state and b) what actions are needed to (accurately) propagate that state in time? 

Take, for example, at a given time, observations of the the wall and plasma facing components in comparison to observations of the core. If we observe tungsten in the edge via spectroscopic measurements\cite{Neu1997ObservationsPlasmas}, this accumulation is not immediately seen in the core. Thus an open question is how to reconcile the differing time scales of differing order phenomena in tokamaks?   

A limitation to our model is that it is data-driven and generative. Outside of constraining the output of the decoder to $\mathcal{R}^+$, we do not enforce the model to predict `physically valid' plasmas. It is of interest, then, to look into how we may constrain the representation to be physically valid.

Future work would explore including MHD stability and instability information into the state. Such a model could provide time-of-flight information on whether a plasma crosses a stability threshold, and if so, possibly what instability may be triggered.

\begin{acknowledgments}
This work has been carried out within the framework of the EUROfusion Consortium, funded by the European Union via the Euratom Research and Training Programme (Grant Agreement No 101052200 — EUROfusion). Views and opinions expressed are however those of the author(s) only and do not necessarily reflect those of the European Union or the European Commission. Neither the European Union nor the European Commission can be held responsible for them.

This work made use of the Finnish CSC computing infrastructure under project \#2005083.  

The work of A.E.J and A.K. was partially supported by the Research Council of Finland grant no. 355460.

The authors would like extend thanks to Professor Satoshi Hamaguchi for hosting the ICDDPS-4 conference, where this work was presented as a oral contribution. 

A.K. would like to thank Ms. Green for fruitful discussions and artistic inspiration throughout the progress of this work. 

A.K. would like to thank Ivan Zaitsev and Kostantinos Papadakis for energetic conversations surrounding the development of this work. 

The authors would also like to thank the reviewers for improving the content of the paper. Additionally, the authors would like to thank Jakub Tomczak for introducing the chilling concept of verifiability to us.
\end{acknowledgments}

\section*{Data Availability Statement}

The experimental data used for training the deep learning models in this work is stored at the data storage facilities of ASDEX-Upgrade and the authors do not have permission to make this data publicly available. However, the training data preparation routines will be provided on request such that anyone with access to the data can regenerate the training dataset. The Python codes encompassing the deep learning algorithms are available in GitHub at \url{https://github.com/DIGIfusion/latent-state-modeling}.

\bibliography{references_formatted}

\begin{thebibliography}{22}%
\makeatletter
\providecommand \@ifxundefined [1]{%
 \@ifx{#1\undefined}
}%
\providecommand \@ifnum [1]{%
 \ifnum #1\expandafter \@firstoftwo
 \else \expandafter \@secondoftwo
 \fi
}%
\providecommand \@ifx [1]{%
 \ifx #1\expandafter \@firstoftwo
 \else \expandafter \@secondoftwo
 \fi
}%
\providecommand \natexlab [1]{#1}%
\providecommand \enquote  [1]{``#1''}%
\providecommand \bibnamefont  [1]{#1}%
\providecommand \bibfnamefont [1]{#1}%
\providecommand \citenamefont [1]{#1}%
\providecommand \href@noop [0]{\@secondoftwo}%
\providecommand \href [0]{\begingroup \@sanitize@url \@href}%
\providecommand \@href[1]{\@@startlink{#1}\@@href}%
\providecommand \@@href[1]{\endgroup#1\@@endlink}%
\providecommand \@sanitize@url [0]{\catcode `\\12\catcode `\$12\catcode
  `\&12\catcode `\#12\catcode `\^12\catcode `\_12\catcode `\%12\relax}%
\providecommand \@@startlink[1]{}%
\providecommand \@@endlink[0]{}%
\providecommand \url  [0]{\begingroup\@sanitize@url \@url }%
\providecommand \@url [1]{\endgroup\@href {#1}{\urlprefix }}%
\providecommand \urlprefix  [0]{URL }%
\providecommand \Eprint [0]{\href }%
\providecommand \doibase [0]{http://dx.doi.org/}%
\providecommand \selectlanguage [0]{\@gobble}%
\providecommand \bibinfo  [0]{\@secondoftwo}%
\providecommand \bibfield  [0]{\@secondoftwo}%
\providecommand \translation [1]{[#1]}%
\providecommand \BibitemOpen [0]{}%
\providecommand \bibitemStop [0]{}%
\providecommand \bibitemNoStop [0]{.\EOS\space}%
\providecommand \EOS [0]{\spacefactor3000\relax}%
\providecommand \BibitemShut  [1]{\csname bibitem#1\endcsname}%
\let\auto@bib@innerbib\@empty
\bibitem [{\citenamefont {Lesort}\ \emph {et~al.}(2018)\citenamefont {Lesort},
  \citenamefont {D{\'{i}}az-Rodr{\'{i}}guez}, \citenamefont {Goudou},\ and\
  \citenamefont {Filliat}}]{Lesort2018StateOverview}%
  \BibitemOpen
  \bibfield  {author} {\bibinfo {author} {\bibfnamefont {T.}~\bibnamefont
  {Lesort}}, \bibinfo {author} {\bibfnamefont {N.}~\bibnamefont
  {D{\'{i}}az-Rodr{\'{i}}guez}}, \bibinfo {author} {\bibfnamefont {J.~F.}\
  \bibnamefont {Goudou}}, \ and\ \bibinfo {author} {\bibfnamefont
  {D.}~\bibnamefont {Filliat}},\ }\bibfield  {title} {\enquote {\bibinfo
  {title} {{State representation learning for control: An overview}},}\ }\href
  {\doibase 10.1016/j.neunet.2018.07.006} {\bibfield  {journal} {\bibinfo
  {journal} {Neural Networks}\ }\textbf {\bibinfo {volume} {108}},\ \bibinfo
  {pages} {379--392} (\bibinfo {year} {2018})}\BibitemShut {NoStop}%
\bibitem [{\citenamefont {Hafner}\ \emph {et~al.}(2023)\citenamefont {Hafner},
  \citenamefont {Pasukonis}, \citenamefont {Ba},\ and\ \citenamefont
  {Lillicrap}}]{Hafner2023MasteringReturn}%
  \BibitemOpen
  \bibfield  {author} {\bibinfo {author} {\bibfnamefont {D.}~\bibnamefont
  {Hafner}}, \bibinfo {author} {\bibfnamefont {J.}~\bibnamefont {Pasukonis}},
  \bibinfo {author} {\bibfnamefont {J.}~\bibnamefont {Ba}}, \ and\ \bibinfo
  {author} {\bibfnamefont {T.}~\bibnamefont {Lillicrap}},\ }\bibfield  {title}
  {\enquote {\bibinfo {title} {Mastering diverse domains through world
  models},}\ }\href@noop {} {\bibfield  {journal} {\bibinfo  {journal} {arXiv
  preprint arXiv:2301.04104}\ } (\bibinfo {year} {2023})}\BibitemShut {NoStop}%
\bibitem [{\citenamefont {Abbate}\ \emph {et~al.}(2023)\citenamefont {Abbate},
  \citenamefont {Conlin}, \citenamefont {Shousha}, \citenamefont {Erickson},\
  and\ \citenamefont {Kolemen}}]{abbate_2023}%
  \BibitemOpen
  \bibfield  {author} {\bibinfo {author} {\bibfnamefont {J.}~\bibnamefont
  {Abbate}}, \bibinfo {author} {\bibfnamefont {R.}~\bibnamefont {Conlin}},
  \bibinfo {author} {\bibfnamefont {R.}~\bibnamefont {Shousha}}, \bibinfo
  {author} {\bibfnamefont {K.}~\bibnamefont {Erickson}}, \ and\ \bibinfo
  {author} {\bibfnamefont {E.}~\bibnamefont {Kolemen}},\ }\bibfield  {title}
  {\enquote {\bibinfo {title} {A general infrastructure for data-driven control
  design and implementation in tokamaks},}\ }\href {\doibase
  10.1017/S0022377822001040} {\bibfield  {journal} {\bibinfo  {journal}
  {Journal of Plasma Physics}\ }\textbf {\bibinfo {volume} {89}},\ \bibinfo
  {pages} {895890102} (\bibinfo {year} {2023})}\BibitemShut {NoStop}%
\bibitem [{\citenamefont {Kit}\ \emph {et~al.}(2023)\citenamefont {Kit},
  \citenamefont {J{\"{a}}rvinen}, \citenamefont {Wiesen}, \citenamefont
  {Poels},\ and\ \citenamefont {Frassinetti}}]{Kit2023DevelopingJET}%
  \BibitemOpen
  \bibfield  {author} {\bibinfo {author} {\bibfnamefont {A.}~\bibnamefont
  {Kit}}, \bibinfo {author} {\bibfnamefont {A.}~\bibnamefont {J{\"{a}}rvinen}},
  \bibinfo {author} {\bibfnamefont {S.}~\bibnamefont {Wiesen}}, \bibinfo
  {author} {\bibfnamefont {Y.}~\bibnamefont {Poels}}, \ and\ \bibinfo {author}
  {\bibfnamefont {L.}~\bibnamefont {Frassinetti}},\ }\bibfield  {title}
  {\enquote {\bibinfo {title} {{Developing deep learning algorithms for
  inferring upstream separatrix density at JET}},}\ }\href {\doibase
  10.1016/j.nme.2022.101347} {\bibfield  {journal} {\bibinfo  {journal}
  {Nuclear Materials and Energy}\ }\textbf {\bibinfo {volume} {34}} (\bibinfo
  {year} {2023}),\ 10.1016/j.nme.2022.101347}\BibitemShut {NoStop}%
\bibitem [{\citenamefont {Zhu}\ \emph {et~al.}(2022)\citenamefont {Zhu},
  \citenamefont {Zhao}, \citenamefont {Bhatia}, \citenamefont {Xu},
  \citenamefont {Bremer}, \citenamefont {Meyer}, \citenamefont {Li},\ and\
  \citenamefont {Rognlien}}]{Zhu2022Data-drivenPrediction}%
  \BibitemOpen
  \bibfield  {author} {\bibinfo {author} {\bibfnamefont {B.}~\bibnamefont
  {Zhu}}, \bibinfo {author} {\bibfnamefont {M.}~\bibnamefont {Zhao}}, \bibinfo
  {author} {\bibfnamefont {H.}~\bibnamefont {Bhatia}}, \bibinfo {author}
  {\bibfnamefont {X.-q.}\ \bibnamefont {Xu}}, \bibinfo {author} {\bibfnamefont
  {P.-T.}\ \bibnamefont {Bremer}}, \bibinfo {author} {\bibfnamefont
  {W.}~\bibnamefont {Meyer}}, \bibinfo {author} {\bibfnamefont
  {N.}~\bibnamefont {Li}}, \ and\ \bibinfo {author} {\bibfnamefont
  {T.}~\bibnamefont {Rognlien}},\ }\bibfield  {title} {\enquote {\bibinfo
  {title} {{Data-driven model for divertor plasma detachment prediction}},}\
  }\href@noop {} {\bibfield  {journal} {\bibinfo  {journal} {Journal of of
  Plasma Physics}\ } (\bibinfo {year} {2022})}\BibitemShut {NoStop}%
\bibitem [{\citenamefont {Joy}\ \emph {et~al.}(2021)\citenamefont {Joy},
  \citenamefont {Schmon}, \citenamefont {Torr}, \citenamefont {N},\ and\
  \citenamefont {Rainforth}}]{JoyCAPTURINGVAEs}%
  \BibitemOpen
  \bibfield  {author} {\bibinfo {author} {\bibfnamefont {T.}~\bibnamefont
  {Joy}}, \bibinfo {author} {\bibfnamefont {S.}~\bibnamefont {Schmon}},
  \bibinfo {author} {\bibfnamefont {P.}~\bibnamefont {Torr}}, \bibinfo {author}
  {\bibfnamefont {S.}~\bibnamefont {N}}, \ and\ \bibinfo {author}
  {\bibfnamefont {T.}~\bibnamefont {Rainforth}},\ }\bibfield  {title} {\enquote
  {\bibinfo {title} {Capturing label characteristics in {\{}vae{\}}s},}\ }in\
  \href {https://openreview.net/forum?id=wQRlSUZ5V7B} {\emph {\bibinfo
  {booktitle} {International Conference on Learning Representations}}}\
  (\bibinfo {year} {2021})\BibitemShut {NoStop}%
\bibitem [{\citenamefont {Kingma}\ and\ \citenamefont
  {Welling}(2014)}]{Kingma2014Auto-encodingBayes}%
  \BibitemOpen
  \bibfield  {author} {\bibinfo {author} {\bibfnamefont {D.~P.}\ \bibnamefont
  {Kingma}}\ and\ \bibinfo {author} {\bibfnamefont {M.}~\bibnamefont
  {Welling}},\ }\bibfield  {title} {\enquote {\bibinfo {title} {{Auto-encoding
  variational bayes}},}\ }\href@noop {} {\bibfield  {journal} {\bibinfo
  {journal} {2nd International Conference on Learning Representations, ICLR
  2014 - Conference Track Proceedings}\ ,\ \bibinfo {pages} {8--9}} (\bibinfo
  {year} {2014})}\BibitemShut {NoStop}%
\bibitem [{\citenamefont {Fischer}\ \emph {et~al.}(2010)\citenamefont
  {Fischer}, \citenamefont {Fuchs}, \citenamefont {Kurzan}, \citenamefont
  {Suttrop},\ and\ \citenamefont {Wolfrum}}]{Fischer2010IntegratedUpgrade}%
  \BibitemOpen
  \bibfield  {author} {\bibinfo {author} {\bibfnamefont {R.}~\bibnamefont
  {Fischer}}, \bibinfo {author} {\bibfnamefont {C.~J.}\ \bibnamefont {Fuchs}},
  \bibinfo {author} {\bibfnamefont {B.}~\bibnamefont {Kurzan}}, \bibinfo
  {author} {\bibfnamefont {W.}~\bibnamefont {Suttrop}}, \ and\ \bibinfo
  {author} {\bibfnamefont {E.}~\bibnamefont {Wolfrum}},\ }\bibfield  {title}
  {\enquote {\bibinfo {title} {{Integrated data analysis of profile diagnostics
  at ASDEX upgrade}},}\ }\href {\doibase 10.13182/FST10-110} {\bibfield
  {journal} {\bibinfo  {journal} {Fusion Science and Technology}\ }\textbf
  {\bibinfo {volume} {58}},\ \bibinfo {pages} {675--684} (\bibinfo {year}
  {2010})}\BibitemShut {NoStop}%
\bibitem [{\citenamefont {Takizuka}\ and\ \citenamefont
  {Martin}(2008)}]{Takizuka2008PowerITER}%
  \BibitemOpen
  \bibfield  {author} {\bibinfo {author} {\bibfnamefont {T.}~\bibnamefont
  {Takizuka}}\ and\ \bibinfo {author} {\bibfnamefont {Y.~R.}\ \bibnamefont
  {Martin}},\ }\bibfield  {title} {\enquote {\bibinfo {title} {{Power
  requirement for accessing the H-mode in ITER}},}\ }\href {\doibase
  10.1088/1742-6596/123/1/012033} {\bibfield  {journal} {\bibinfo  {journal}
  {Journal of Physics: Conference Series OPEN ACCESS}\ }\textbf {\bibinfo
  {volume} {123}},\ \bibinfo {pages} {12033} (\bibinfo {year}
  {2008})}\BibitemShut {NoStop}%
\bibitem [{\citenamefont {Ha}\ and\ \citenamefont
  {Schmidhuber}(2018)}]{HaWorldModels}%
  \BibitemOpen
  \bibfield  {author} {\bibinfo {author} {\bibfnamefont {D.}~\bibnamefont
  {Ha}}\ and\ \bibinfo {author} {\bibfnamefont {J.}~\bibnamefont
  {Schmidhuber}},\ }\bibfield  {title} {\enquote {\bibinfo {title} {World
  models},}\ }\href {\doibase 10.5281/ZENODO.1207631} {\  (\bibinfo {year}
  {2018}),\ 10.5281/ZENODO.1207631}\BibitemShut {NoStop}%
\bibitem [{\citenamefont {Hafner}\ \emph {et~al.}(2019)\citenamefont {Hafner},
  \citenamefont {Lillicrap}, \citenamefont {Fischer}, \citenamefont {Villegas},
  \citenamefont {Ha}, \citenamefont {Lee},\ and\ \citenamefont
  {Davidson}}]{HafnerLearningPixels}%
  \BibitemOpen
  \bibfield  {author} {\bibinfo {author} {\bibfnamefont {D.}~\bibnamefont
  {Hafner}}, \bibinfo {author} {\bibfnamefont {T.}~\bibnamefont {Lillicrap}},
  \bibinfo {author} {\bibfnamefont {I.}~\bibnamefont {Fischer}}, \bibinfo
  {author} {\bibfnamefont {R.}~\bibnamefont {Villegas}}, \bibinfo {author}
  {\bibfnamefont {D.}~\bibnamefont {Ha}}, \bibinfo {author} {\bibfnamefont
  {H.}~\bibnamefont {Lee}}, \ and\ \bibinfo {author} {\bibfnamefont
  {J.}~\bibnamefont {Davidson}},\ }\bibfield  {title} {\enquote {\bibinfo
  {title} {Learning latent dynamics for planning from pixels},}\ }in\
  \href@noop {} {\emph {\bibinfo {booktitle} {International Conference on
  Machine Learning}}}\ (\bibinfo {year} {2019})\ pp.\ \bibinfo {pages}
  {2555--2565}\BibitemShut {NoStop}%
\bibitem [{\citenamefont {Brandstetter}, \citenamefont {Worrall},\ and\
  \citenamefont {Welling}(2023)}]{BrandstetterMESSAGESOLVERS}%
  \BibitemOpen
  \bibfield  {author} {\bibinfo {author} {\bibfnamefont {J.}~\bibnamefont
  {Brandstetter}}, \bibinfo {author} {\bibfnamefont {D.}~\bibnamefont
  {Worrall}}, \ and\ \bibinfo {author} {\bibfnamefont {M.}~\bibnamefont
  {Welling}},\ }\href@noop {} {\enquote {\bibinfo {title} {Message passing
  neural pde solvers},}\ } (\bibinfo {year} {2023}),\ \Eprint
  {http://arxiv.org/abs/2202.03376} {arXiv:2202.03376 [cs.LG]} \BibitemShut
  {NoStop}%
\bibitem [{\citenamefont {Kingma}\ and\ \citenamefont
  {Ba}(2014)}]{Kingma2014Adam:Optimization}%
  \BibitemOpen
  \bibfield  {author} {\bibinfo {author} {\bibfnamefont {D.~P.}\ \bibnamefont
  {Kingma}}\ and\ \bibinfo {author} {\bibfnamefont {J.~L.}\ \bibnamefont
  {Ba}},\ }\bibfield  {title} {\enquote {\bibinfo {title} {{Adam: A Method for
  Stochastic Optimization}},}\ }\href {https://arxiv.org/abs/1412.6980v9}
  {\bibfield  {journal} {\bibinfo  {journal} {3rd International Conference on
  Learning Representations, ICLR 2015 - Conference Track Proceedings}\ }
  (\bibinfo {year} {2014})}\BibitemShut {NoStop}%
\bibitem [{\citenamefont {Matos}\ \emph {et~al.}(2021)\citenamefont {Matos},
  \citenamefont {Menkovski}, \citenamefont {Pau}, \citenamefont {Marceca},
  \citenamefont {Jenko},\ and\ \citenamefont {the
  TCV~Team}}]{2021PlasmaAttention}%
  \BibitemOpen
  \bibfield  {author} {\bibinfo {author} {\bibfnamefont {F.}~\bibnamefont
  {Matos}}, \bibinfo {author} {\bibfnamefont {V.}~\bibnamefont {Menkovski}},
  \bibinfo {author} {\bibfnamefont {A.}~\bibnamefont {Pau}}, \bibinfo {author}
  {\bibfnamefont {G.}~\bibnamefont {Marceca}}, \bibinfo {author} {\bibfnamefont
  {F.}~\bibnamefont {Jenko}}, \ and\ \bibinfo {author} {\bibnamefont {the
  TCV~Team}},\ }\bibfield  {title} {\enquote {\bibinfo {title} {Plasma
  confinement mode classification using a sequence-to-sequence neural network
  with attention},}\ }\href {\doibase 10.1088/1741-4326/abe370} {\bibfield
  {journal} {\bibinfo  {journal} {Nuclear Fusion}\ }\textbf {\bibinfo {volume}
  {61}},\ \bibinfo {pages} {046019} (\bibinfo {year} {2021})}\BibitemShut
  {NoStop}%
\bibitem [{\citenamefont {Neu}\ \emph {et~al.}(2013)\citenamefont {Neu},
  \citenamefont {Kallenbach}, \citenamefont {Balden}, \citenamefont {Bobkov},
  \citenamefont {Coenen}, \citenamefont {Drube}, \citenamefont {Dux},
  \citenamefont {Greuner}, \citenamefont {Herrmann}, \citenamefont {Hobirk},
  \citenamefont {H{\"{o}}hnle}, \citenamefont {Krieger}, \citenamefont
  {Ko{\v{c}}an}, \citenamefont {Lang}, \citenamefont {Lunt}, \citenamefont
  {Maier}, \citenamefont {Mayer}, \citenamefont {M{\"{u}}ller}, \citenamefont
  {Potzel}, \citenamefont {P{\"{u}}tterich}, \citenamefont {Rapp},
  \citenamefont {Rohde}, \citenamefont {Ryter}, \citenamefont {Schneider},
  \citenamefont {Schweinzer}, \citenamefont {Sertoli}, \citenamefont {Stober},
  \citenamefont {Suttrop}, \citenamefont {Sugiyama}, \citenamefont
  {Van~Rooij},\ and\ \citenamefont {Wischmeier}}]{Neu2013OverviewUpgrade}%
  \BibitemOpen
  \bibfield  {author} {\bibinfo {author} {\bibfnamefont {R.}~\bibnamefont
  {Neu}}, \bibinfo {author} {\bibfnamefont {A.}~\bibnamefont {Kallenbach}},
  \bibinfo {author} {\bibfnamefont {M.}~\bibnamefont {Balden}}, \bibinfo
  {author} {\bibfnamefont {V.}~\bibnamefont {Bobkov}}, \bibinfo {author}
  {\bibfnamefont {J.~W.}\ \bibnamefont {Coenen}}, \bibinfo {author}
  {\bibfnamefont {R.}~\bibnamefont {Drube}}, \bibinfo {author} {\bibfnamefont
  {R.}~\bibnamefont {Dux}}, \bibinfo {author} {\bibfnamefont {H.}~\bibnamefont
  {Greuner}}, \bibinfo {author} {\bibfnamefont {A.}~\bibnamefont {Herrmann}},
  \bibinfo {author} {\bibfnamefont {J.}~\bibnamefont {Hobirk}}, \bibinfo
  {author} {\bibfnamefont {H.}~\bibnamefont {H{\"{o}}hnle}}, \bibinfo {author}
  {\bibfnamefont {K.}~\bibnamefont {Krieger}}, \bibinfo {author} {\bibfnamefont
  {M.}~\bibnamefont {Ko{\v{c}}an}}, \bibinfo {author} {\bibfnamefont
  {P.}~\bibnamefont {Lang}}, \bibinfo {author} {\bibfnamefont {T.}~\bibnamefont
  {Lunt}}, \bibinfo {author} {\bibfnamefont {H.}~\bibnamefont {Maier}},
  \bibinfo {author} {\bibfnamefont {M.}~\bibnamefont {Mayer}}, \bibinfo
  {author} {\bibfnamefont {H.~W.}\ \bibnamefont {M{\"{u}}ller}}, \bibinfo
  {author} {\bibfnamefont {S.}~\bibnamefont {Potzel}}, \bibinfo {author}
  {\bibfnamefont {T.}~\bibnamefont {P{\"{u}}tterich}}, \bibinfo {author}
  {\bibfnamefont {J.}~\bibnamefont {Rapp}}, \bibinfo {author} {\bibfnamefont
  {V.}~\bibnamefont {Rohde}}, \bibinfo {author} {\bibfnamefont
  {F.}~\bibnamefont {Ryter}}, \bibinfo {author} {\bibfnamefont {P.~A.}\
  \bibnamefont {Schneider}}, \bibinfo {author} {\bibfnamefont {J.}~\bibnamefont
  {Schweinzer}}, \bibinfo {author} {\bibfnamefont {M.}~\bibnamefont {Sertoli}},
  \bibinfo {author} {\bibfnamefont {J.}~\bibnamefont {Stober}}, \bibinfo
  {author} {\bibfnamefont {W.}~\bibnamefont {Suttrop}}, \bibinfo {author}
  {\bibfnamefont {K.}~\bibnamefont {Sugiyama}}, \bibinfo {author}
  {\bibfnamefont {G.}~\bibnamefont {Van~Rooij}}, \ and\ \bibinfo {author}
  {\bibfnamefont {M.}~\bibnamefont {Wischmeier}},\ }\bibfield  {title}
  {\enquote {\bibinfo {title} {{Overview on plasma operation with a full
  tungsten wall in ASDEX Upgrade}},}\ }\href {\doibase
  10.1016/J.JNUCMAT.2013.01.006} {\bibfield  {journal} {\bibinfo  {journal}
  {Journal of Nuclear Materials}\ }\textbf {\bibinfo {volume} {438}},\ \bibinfo
  {pages} {S34--S41} (\bibinfo {year} {2013})}\BibitemShut {NoStop}%
\bibitem [{\citenamefont {Karl}\ \emph {et~al.}(2017)\citenamefont {Karl},
  \citenamefont {Soelch}, \citenamefont {Bayer},\ and\ \citenamefont {van~der
  Smagt}}]{KarlDEEPDATA}%
  \BibitemOpen
  \bibfield  {author} {\bibinfo {author} {\bibfnamefont {M.}~\bibnamefont
  {Karl}}, \bibinfo {author} {\bibfnamefont {M.}~\bibnamefont {Soelch}},
  \bibinfo {author} {\bibfnamefont {J.}~\bibnamefont {Bayer}}, \ and\ \bibinfo
  {author} {\bibfnamefont {P.}~\bibnamefont {van~der Smagt}},\ }\href@noop {}
  {\enquote {\bibinfo {title} {Deep variational bayes filters: Unsupervised
  learning of state space models from raw data},}\ } (\bibinfo {year} {2017}),\
  \Eprint {http://arxiv.org/abs/1605.06432} {arXiv:1605.06432 [stat.ML]}
  \BibitemShut {NoStop}%
\bibitem [{\citenamefont {Becker-Ehmck}, \citenamefont {Peters},\ and\
  \citenamefont {van~der Smagt}(2019)}]{Becker-EhmckSwitchingFiltering}%
  \BibitemOpen
  \bibfield  {author} {\bibinfo {author} {\bibfnamefont {P.}~\bibnamefont
  {Becker-Ehmck}}, \bibinfo {author} {\bibfnamefont {J.}~\bibnamefont
  {Peters}}, \ and\ \bibinfo {author} {\bibfnamefont {P.}~\bibnamefont {van~der
  Smagt}},\ }\href@noop {} {\enquote {\bibinfo {title} {Switching linear
  dynamics for variational bayes filtering},}\ } (\bibinfo {year} {2019}),\
  \Eprint {http://arxiv.org/abs/1905.12434} {arXiv:1905.12434 [stat.ML]}
  \BibitemShut {NoStop}%
\bibitem [{\citenamefont {Lusch}, \citenamefont {Kutz},\ and\ \citenamefont
  {Brunton}(2018)}]{Lusch2018DeepDynamics}%
  \BibitemOpen
  \bibfield  {author} {\bibinfo {author} {\bibfnamefont {B.}~\bibnamefont
  {Lusch}}, \bibinfo {author} {\bibfnamefont {J.~N.}\ \bibnamefont {Kutz}}, \
  and\ \bibinfo {author} {\bibfnamefont {S.~L.}\ \bibnamefont {Brunton}},\
  }\bibfield  {title} {\enquote {\bibinfo {title} {{Deep learning for universal
  linear embeddings of nonlinear dynamics}},}\ }\href {\doibase
  10.1038/s41467-018-07210-0} {\bibfield  {journal} {\bibinfo  {journal}
  {Nature Communications 2018 9:1}\ }\textbf {\bibinfo {volume} {9}},\ \bibinfo
  {pages} {1--10} (\bibinfo {year} {2018})}\BibitemShut {NoStop}%
\bibitem [{\citenamefont {Ryzhikov}, \citenamefont {Hushchyn},\ and\
  \citenamefont {Derkach}(2022)}]{RyzhikovLatentDetection}%
  \BibitemOpen
  \bibfield  {author} {\bibinfo {author} {\bibfnamefont {A.}~\bibnamefont
  {Ryzhikov}}, \bibinfo {author} {\bibfnamefont {M.}~\bibnamefont {Hushchyn}},
  \ and\ \bibinfo {author} {\bibfnamefont {D.}~\bibnamefont {Derkach}},\
  }\href@noop {} {\enquote {\bibinfo {title} {Latent neural stochastic
  differential equations for change point detection},}\ } (\bibinfo {year}
  {2022}),\ \Eprint {http://arxiv.org/abs/2208.10317} {arXiv:2208.10317
  [cs.LG]} \BibitemShut {NoStop}%
\bibitem [{\citenamefont {Balsells-Rodas}, \citenamefont {Wang},\ and\
  \citenamefont {Li}()}]{balsells-rodas_identifiability_2023}%
  \BibitemOpen
  \bibfield  {author} {\bibinfo {author} {\bibfnamefont {C.}~\bibnamefont
  {Balsells-Rodas}}, \bibinfo {author} {\bibfnamefont {Y.}~\bibnamefont
  {Wang}}, \ and\ \bibinfo {author} {\bibfnamefont {Y.}~\bibnamefont {Li}},\
  }\href {http://arxiv.org/abs/2305.15925} {\enquote {\bibinfo {title} {On the
  identifiability of markov switching models},}\ }\Eprint
  {http://arxiv.org/abs/2305.15925 [cs, stat]} {2305.15925 [cs, stat]}
  \BibitemShut {NoStop}%
\bibitem [{\citenamefont {Ilse}\ \emph {et~al.}(2019)\citenamefont {Ilse},
  \citenamefont {Tomczak}, \citenamefont {Louizos},\ and\ \citenamefont
  {Welling}}]{ilse2019diva}%
  \BibitemOpen
  \bibfield  {author} {\bibinfo {author} {\bibfnamefont {M.}~\bibnamefont
  {Ilse}}, \bibinfo {author} {\bibfnamefont {J.~M.}\ \bibnamefont {Tomczak}},
  \bibinfo {author} {\bibfnamefont {C.}~\bibnamefont {Louizos}}, \ and\
  \bibinfo {author} {\bibfnamefont {M.}~\bibnamefont {Welling}},\ }\href@noop
  {} {\enquote {\bibinfo {title} {Diva: Domain invariant variational
  autoencoders},}\ } (\bibinfo {year} {2019}),\ \Eprint
  {http://arxiv.org/abs/1905.10427} {arXiv:1905.10427 [stat.ML]} \BibitemShut
  {NoStop}%
\bibitem [{\citenamefont {Neu}\ \emph {et~al.}(1997)\citenamefont {Neu},
  \citenamefont {Fournier}, \citenamefont {Schlögl},\ and\ \citenamefont
  {Rice}}]{Neu1997ObservationsPlasmas}%
  \BibitemOpen
  \bibfield  {author} {\bibinfo {author} {\bibfnamefont {R.}~\bibnamefont
  {Neu}}, \bibinfo {author} {\bibfnamefont {K.~B.}\ \bibnamefont {Fournier}},
  \bibinfo {author} {\bibfnamefont {D.}~\bibnamefont {Schlögl}}, \ and\
  \bibinfo {author} {\bibfnamefont {J.}~\bibnamefont {Rice}},\ }\bibfield
  {title} {\enquote {\bibinfo {title} {Observations of x-ray spectra from
  highly charged tungsten ions in tokamak plasmas},}\ }\href {\doibase
  10.1088/0953-4075/30/21/036} {\bibfield  {journal} {\bibinfo  {journal}
  {Journal of Physics B: Atomic, Molecular and Optical Physics}\ }\textbf
  {\bibinfo {volume} {30}},\ \bibinfo {pages} {5057} (\bibinfo {year}
  {1997})}\BibitemShut {NoStop}%
\end{thebibliography}%

\end{document}